\documentclass[twocolumn,preprintnumbers]{revtex4}

\usepackage{graphicx}
\usepackage{epsf}
\usepackage{amsmath,amssymb}
\usepackage{color}

\begin{document}

\newcommand{\bvec}{\boldsymbol}
\newcommand{\La}{\Lambda}
\newcommand{\LLa}{{\Lambda\Lambda}}
\newcommand{\LLi}{_{\La}\textrm{Li}}
\newcommand{\LBe}{_{\La}\textrm{Be}}
\newcommand{\LB}{_{\La}\textrm{B}}

\newcommand{\LC}{_{\La}\textrm{C}}
\newcommand{\LO}{^{16}_{\La}\textrm{O}}
\newcommand{\LF}{^{19}_{\La}\textrm{F}}

\newcommand{\CLi}{\textrm{Li}}
\newcommand{\CBe}{\textrm{Be}}
\newcommand{\CB}{\textrm{B}}

\newcommand{\CC}{\textrm{C}}
\newcommand{\CO}{^{15}\textrm{O}}
\newcommand{\CF}{^{18}\textrm{F}}

\newcommand{\CZ}{^{A-1}Z}

\newcommand{\LZ}{^{A}_\La Z}
\newcommand{\LLZ}{^{\ \,A}_{\La\La} Dummy}

\newcommand{\red}[1]{\textcolor{red}{#1}}
\newcommand{\blue}[1]{\textcolor{cyan}{#1}}


\title{Energy spectra in $p$-shell $\Lambda$ hypernuclei and $\LF$ and spin-dependent  $\La N$ interactions}
\author{Yoshiko Kanada-En'yo}
\affiliation{Department of Physics, Kyoto University, Kyoto 606-8502, Japan}
\author{Masahiro Isaka}
\affiliation{Research Center for Nuclear Physics (RCNP), Osaka University, Ibaraki, Osaka, 567-0047, Japan}
\author{Toshio Motoba}
\affiliation{Laboratory of Physics, Osaka Electro-Communication University, Neyagawa 572-8530, Japan}
\affiliation{Yukawa Institute for Theoretical Physics, Kyoto University, Kyoto 606-8502, Japan}
\begin{abstract}
Energy spectra of $0s$-orbit $\La$ states in $p$-shell $\Lambda$ hypernuclei ($\LZ$) and 
those in $\LF$ are studied 
with the microscopic cluster model and antisymmetrized molecular dynamics
using the $G$-matrix effective $\La N$ ($\La NG$) interactions. 
Spin-dependent terms of the ESC08a version of the $\La NG$ interactions are tested and 
phenomenologically tuned to reproduce observed energy spectra 
in $p$-shell $\LZ$. Spin-dependent 
contributions of the $\La N$ interactions to spin-doublet splitting and 
excitation energies are discussed. 
Energy spectra for unobserved excited states in $p$-shell $\LZ$ and $\LF$ are predicted with
the modified $\La NG$ interactions.
\end{abstract}
\maketitle

\section{Introduction}

In this decade, experimental and theoretical 
studies of hypernuclei have remarkably progressed. 
For $\La$ hypernuclei, experimental studies with high-resolution $\gamma$-ray measurements 
have been extensively performed to provide detailed information of 
energy spectra in the $p$-shell region \cite{Hashimoto:2006aw,Tamura:2010zz,
Tamura:2013lwa}. 
The
$\gamma$-ray spectroscopic study 
of $sd$-shell $\LZ$ has just started, and 
the observed spectra in $\LF$ have been reported \cite{Yang:2017lay}. 
The observed energy spectra are useful information for study of
$\Lambda N$ interactions. In particular, energy splittings between the spin-doublet 
$J_>=I+1/2$ and $J_<=I-1/2$ states with the $\La$-spin coupling in parallel 
and anti-parallel to the core nuclear spin $I$ are sensitive probes to figure out 
spin dependences of the $\La N$ interactions in $\La$ hypernuclei. 
Based on compilation of the precise data updated recently,  
it is time to comprehensively understand the energy spectra in $p$-shell $\LZ$ with theoretical study.
Moreover, it is able to test spin dependences of the $\La N$ interactions in comparison of calculated spectra with 
observed data.

Structure studies of $p$-shell $\Lambda$ hypernuclei have been 
performed with various theoretical models such as 
cluster models \cite{Motoba:1984ri,motoba85,Yamada:1985qr,Yu:1986ip,Hiyama:1996gv,Hiyama:1997ub,Hiyama:1999me,Hiyama:2000jd,Hiyama:2002yj,Hiyama:2006xv,Hiyama:2010zzc,Hiyama:2012sq,Cravo:2002jv,Suslov:2004ed,Mohammad:2009zza,Zhang:2012zzg,Funaki:2014fba,Funaki:2017asz}, shell models (SM) \cite{Gal:1971gb,Gal:1972gd,Gal:1978jt,Millener:2008zz,Millener:2010zz,Millener:2012zz}, mean-field and beyond mean-field models \cite{Guleria:2011kk,Vidana:2001rm,Zhou:2007zze,Win:2008vw,Win:2010tq,Lu:2011wy,Mei:2014hya,Mei:2015pca,Mei:2016lce,Schulze:2014oia}, hyper antisymmetrized molecular dynamics (HAMD) model
\cite{Isaka:2011kz,Isaka:2015xda,Homma:2015kia,Isaka:2016apm,Isaka:2017nuc},  and no-core shell model \cite{Wirth:2014apa}, and so on.
Spin-dependent $\La N$ interactions have been discussed in details in systematic studies for various $p$-shell $\LZ$, for instance,  with 
cluster-model \cite{Hiyama:2000jd,Hiyama:2006xv,Hiyama:2010zzc,Hiyama:2012sq} 
and shell-model calculations \cite{Millener:2008zz,Millener:2010zz,Millener:2012zz}. 

In the structure calculations of $\Lambda$ hypernuclei, $YN$ interactions developed based on the 
meson-theoretical models by the Nijmegen group have been widely used. After many trials and 
continuous improvements,  new versions (ESC08 series) 
of the extended-soft-core (ESC) model of 
the $YN$ interactions have been proposed \cite{Rijken:2010zza,Rijken:2010zzb,Yamamoto:2010zzn}.
The spin-independent part of the ESC08(a,b) $\La N$ interaction was 
tested and found to be reasonable in description of  $\La$ binding energies in a wide mass number region
\cite{Yamamoto:2010zzn, Isaka:2016apm,Isaka:2017nuc}.
However, in spin dependences of the ESC08(a,b), 
problems were found in reproducing the observed spin-doublet splitting energies \cite{Yamamoto:2010zzn}.
It is a demanded issue to test the spin dependences of the ESC08 with systematic investigation of energy spectra 
in $p$-shell $\LZ$ and consider possible modification of the spin-dependent $\La N$ interactions 
in comparison with the experimental data.

It is known  from observed $\La$ binding energies that 
the $\La N$ interactions are weak compared with the $NN$ interactions.
Moreover, from observed energy spectra,  the spin-dependent $\La N$ interactions have been found to be 
rather weak compared with the spin-independent $\La N$ interactions, and therefore, they may give perturbative contributions to 
structures of $\LZ$.
It means that the $\La$ particle in $\LZ$, an impurity embedded in nuclear system,  is regarded as a spectator 
probing the $\La N$ interactions. In particular, energy spectra of $(0s)_\La$ states can be a good probe that 
detects rather directly the spin-dependent $\La N$ interactions through the spin-dependent mean-filed potential 
for the $0s$-orbit $\La$ determined by nuclear spin structure in core nuclei.
In order to describe detailed energy spectra and understand properties of the spin-dependent $\La N$ interactions, 
one needs a reliable structure model which can properly describe nuclear structures, particularly, nuclear spin configurations.
Furthermore, for systematic studies of  $p$-shell $\LZ$, 
it is also demanded to describe cluster structures in light-mass $p$-shell nuclei. 
Cluster models can respond to the latter demand, but in general they are not sufficient in describing nuclear 
spin configurations in med-$p$-shell nuclei because cluster breakings are not taken into account in the model.
Shell models are useful to investigate detailed spin configurations, but it is not suitable to deal with 
remarkable clustering as well as nuclear deformations because of limitation of the model space. 
The antisymmetrized molecular dynamics (AMD) model \cite{KanadaEnyo:1995tb,KanadaEnyo:1995ir,AMDsupp,KanadaEn'yo:2012bj} is one of  useful tools
for systematic study of $p$-shell and $sd$-shell nuclei
because it can describe cluster and spin structures in the ground and excited states of 
general nuclei.  A version of the AMD model,  variation after projection
called AMD+VAP,  has been applied to various $p$-shell nuclei including odd-odd nuclei and proved to be successful
in describing nuclear spin properties such as $\mu$ moments, 
$M1$, and $GT$ transitions \cite{KanadaEn'yo:1998rf,KanadaEn'yo:2012bj,Kanada-Enyo:2014qwn,Kanada-Enyo:2015uiy}. 
The HAMD,  which is another version of the AMD applied to 
$\La$ hypernuclei by one of the authors (M. I.)  and his collaborators \cite{Isaka:2011kz,Isaka:2015xda,Homma:2015kia,Isaka:2016apm,Isaka:2017nuc}, is also a 
promising approach for study of spin-dependent $\La N$ interactions though its application is still limited. 

The aim of the present work is to investigate energy spectra in $p$-shell $\LZ$ and discuss
spin dependences of the $\La N$ interactions with microscopic structure model calculations.
In the previous works by one of the authors (Y. K-E.), 
spin-averaged energy spectra of low-lying $(0s)_\La$ states have 
been investigated
by applying the cluster model for core nuclei 
and a single-channel potential model for a $\Lambda$ particle with 
the spin-independent effective $\La N$ interactions \cite{Kanada-Enyo:2017ynk}.
In order to calculate energy spectra
of $\LZ$ with spin-dependent effective $\La N$ interactions, 
here we apply the AMD+VAP model 
in addition to the microscopic cluster model.
The spin-dependent $\La N$ interactions are perturbatively treated in the AMD+VAP calculation.
Comparing the calculated spin-doublet splitting energies with 
observed data in $p$-shell $\LZ$,  we test spin dependences of the 
$G$-matrix effective $\La N$ ($\La NG$) interactions of the ESC08a model 
\cite{Rijken:2010zza,Rijken:2010zzb,Yamamoto:2010zzn}. 
A modification of 
the ESC08a $\La NG$ interaction is proposed by phenomenological tuning of  
the spin-dependent terms to adjust available data of energy spectra. 
Using the modified $\La NG$ interactions, the spin-dependent contributions 
of the $\La N$ interactions to energy spectra are investigated. In addition, 
theoretical spectra for unknown excited states in $p$-shell $\LZ$ and
$\LF$ are predicted.

This paper is organized as follows. In the next section, we explain the framework of the present calculation.
The effective $NN$ and $\Lambda N$
interactions are explained in Sec.~\ref{sec:interactions}. Structure properties of core nuclei $\CZ$ are shown in 
Sec.~\ref{sec:CZ}.  In Sec.~\ref{sec:results}, results for $\LZ$ and the modification of the
spin-dependent $\La N$ interactions are given. 
The paper is summarized in Sec.~\ref{sec:summary}. In
appendixes \ref{app:folding-pot} and \ref{app:rearrangements}, validity of the folding potential model approximation
and spin rearrangement effects are discussed, respectively. 

\section{Framework} \label{sec:framework}

We apply two models to describe structures of core nuclear part. 
One is the microscopic cluster model 
with the generator coordinate method (GCM) \cite{Hill:1952jb,Griffin:1957zza}, which has been used in the previous works
\cite{Kanada-Enyo:2017ynk,Kanada-Enyo:2018pxt}, 
and the other is the AMD+VAP.
The framework of the AMD model is explained, for example, in Ref.~\cite{KanadaEn'yo:2012bj}.
For details of the frameworks, see those papers and references therein.

In the cluster model with the GCM, 
the dynamical inter-cluster motion is taken into account by means of superposition of cluster wave functions
having various inter-cluster distances. 
However,  the cluster model contains 
only a part of intrinsic-spin configurations because it ignores cluster breaking. 
To overcome this problem of the cluster model and 
investigate spin-dependent contributions of the $\La N$  interactions to energy spectra, 
we apply the AMD model. A basis AMD wave function is given by a Slater determinant of single-nucleon 
Gaussian wave functions, in which Gaussian centroids and intrinsic-spin orientations of all nucleons are independently 
treated as variational parameters. The AMD model does not rely on {\it a priori} assumption of clusters and 
can describe the cluster breaking and cluster formation. Compared with the cluster model, the AMD is a
flexible model, in particular,  for intrinsic-spin degrees of freedom.
However, in description of dynamical inter-cluster motion, the present AMD+VAP calculation is more limited
than the cluster model because only a few AMD configurations are superposed.

The $\La N$ interactions contain spin-independent ($V_0$) and spin-dependent  ($V_1$) parts as, 
\begin{eqnarray}
V_{\La N}&=&V_0+V_1.
\end{eqnarray}
The spin-independent part ($V_0$) is dominant and gives leading contributions, whereas the spin-dependent part ($V_1$) is relatively weak. 
In the present calculation, we first consider the spin-independent $\La N$ interaction $V_0$ as the leading part for the mean potential 
of the $0s$-orbit $\La$, 
and then perturbatively take into account the spin-dependent $\La N$  interaction $V_1$.
For $p$-shell $\LZ$, we investigate the leading contributions with cluster and AMD models using only $V_0$ by ignoring 
the spin dependence of the $\La N$ interactions. In order to investigate the spin-dependent contributions from $V_1$,  we apply the AMD model. 
For $\LF$, we investigate the leading and perturbative contributions 
with the microscopic cluster model of $^{16}\textrm{O}+p+n$.


\subsection{Calculations of core nuclei with microscopic structure models}

\subsubsection{Cluster model with GCM for $p$-shell nuclei}
Core nuclei $\CZ$ in $\LZ$ 
are calculated with the microscopic cluster model in the same way as the previous
calculations for $p$-shell 
$\La$ and double-$\La$ hypernuclei in Refs.~\cite{Kanada-Enyo:2017ynk,Kanada-Enyo:2018pxt}.
In the model, microscopic $A_N$-nucleon wave functions are expressed by the Brink-Bloch cluster wave functions \cite{Brink66} 
and superposed by means of the GCM. Here, $A_N=A-1$ is the mass number of core nuclei.
The cluster wave functions of 
$\alpha+d$, $\alpha+t$, $2\alpha$, $2\alpha+n$, $2\alpha+p$, $2\alpha+n^2$, $2\alpha+d$, $2\alpha+t$, $2\alpha+h$, $3\alpha$, and 
$3\alpha+h$ cluster wave functions are adopted for $^6\textrm{Li}$,  $^7\textrm{Li}$, $^8\textrm{Be}$,  $^{9}\textrm{Be}$,   $^{9}\textrm{B}$, 
 $^{10}\textrm{Be}$, $^{10}\textrm{B}$,  $^{11}\textrm{B}$,  $^{11}\textrm{C}$, $^{12}\textrm{C}$, and  $\CO$
systems, respectively.
$d$, $n^2$, $t$, $h$, and $\alpha$ clusters are written by harmonic oscillator (h.o.) 
$0s$ configurations. 
For $^{10}$Be, $^{11}\textrm{B}$, $^{11}\textrm{C}$, and  $^{12}\textrm{C}$, 
additional configurations are included in the model space to take into account cluster breaking components
as explained in Ref.~\cite{Kanada-Enyo:2018pxt}.
Namely, 
the $^6\textrm{He}+\alpha$ wave functions \cite{Kanada-Enyo:2016jnq} are 
added to the $2\alpha+n^2$ wave functions for  $^{10}$Be,  and 
the $p_{3/2}$ configurations are added 
to the $2\alpha+t$, $2\alpha+h$, and $3\alpha$ cluster wave functions for  $^{11}\textrm{B}$, $^{11}\textrm{C}$ and  $^{12}\textrm{C}$
\cite{Kanada-Enyo:2017ynk,Suhara:2014wua}. 
Those cluster models (with and without additional configurations) 
are denoted by the label ``CL'' in the paper. 
The h.o. width parameter $\nu$ of clusters is  commonly chosen as 
 $\nu=0.235$ fm$^{-2}$ for $A_N \le 12$ nuclei. For $\CO$,  the value $\nu=0.16$ fm$^{-2}$  which reproduces the 
nuclear size of the $p$-closed $^{16}\textrm{O}$ is used.

The Brink-Bloch cluster wave function
for a $A_N$-nucleon system consisting of $C_1,\ldots,C_k$ clusters is
denoted as $\Phi_\textrm{BB}(\bvec{S}_1,\ldots, \bvec{S}_k)$ with the 
parameters $\bvec{S}_j$ ($j=1,\ldots,k$)  of cluster center positions. 
$k$ is the number of clusters. 
To take into account inter-cluster motion,  the GCM is applied to the angular-momentum and parity projected  
Brink-Bloch cluster wave functions with respect to the generator coordinates $\bvec{S}_j$.
The wave function $\Psi_N(I^\pi_n)$  for the nuclear angular momentum and parity 
$I^\pi_n$ state is given by a linear combination of the 
Brink-Bloch wave functions with various configurations of $\{\bvec{S}_1,\ldots, \bvec{S}_k\}$ as 
\begin{equation}\label{eq:gcm-wf}
\Psi_N(I^\pi_n)=\sum_{\bvec{S}_1,\ldots, \bvec{S}_k} \sum_{K} c^{I^\pi_n}_{\bvec{S}_1,\ldots, \bvec{S}_k,K}
P^{I\pi}_{MK} \Phi_\textrm{BB}(\bvec{S}_1,\ldots, \bvec{S}_k),
\end{equation}
where $P^{I\pi}_{MK}$ is the angular momentum and parity projection operator. The coefficients 
$c^{I^\pi_n}_{\bvec{S}_1,\ldots, \bvec{S}_k,K}$ are determined by solving Griffin-Hill-Wheeler equations 
\cite{Hill:1952jb,Griffin:1957zza}, which is equivalent to the diagonalization of the Hamiltonian and norm matrices,
\begin{eqnarray}
&&\langle  \Phi_\textrm{BB}(\bvec{S}_1,\ldots, \bvec{S}_k) P^{I\pi}_{MK}| H_N|
P^{I\pi}_{MK'} \Phi_\textrm{BB}(\bvec{S}'_1,\ldots, \bvec{S}'_k) \rangle,\nonumber\\ 
&&\langle  \Phi_\textrm{BB}(\bvec{S}_1,\ldots, \bvec{S}_k) P^{I\pi}_{MK}|
P^{I\pi}_{MK'} \Phi_\textrm{BB}(\bvec{S}'_1,\ldots, \bvec{S}'_k) \rangle.\nonumber
\end{eqnarray}
Here $H_N$ is the Hamiltonian of the nuclear part described later.

For the $\alpha+d$ and $2\alpha$ wave functions, $\bvec{S}_1$ and  $\bvec{S}_2$  are chosen to be 
$\bvec{S}_1-\bvec{S}_2=(0,0,d)$ with the generator coordinate $d=\{1,2,\cdots,15$ fm\}
of the inter-cluster distance. 
For the $\alpha+t$ wave functions,  $d=\{1,2,\cdots,8$ fm\} are adopted to obtain a bound state solution for 
the resonance state $^7\textrm{Li}(7/2^-_1)$ corresponding to a bound state approximation. 
For configurations of $2\alpha+n^2$, $2\alpha+d$, $2\alpha+t$, $2\alpha+h$, and $3\alpha$
cluster wave functions, $\bvec{S}_{1,2,3}$ are chosen to be 
\begin{eqnarray}
&&\bvec{S}_1-\bvec{S}_2=(0,0,d), \\
&&\bvec{S}_3-\frac{A_2\bvec{S}_1+A_1\bvec{S}_2}{A_1+A_2}
=(r\sin \theta,0,r \cos \theta),
\end{eqnarray}
with $d=\{1.2,2.2,\ldots,4.2$ fm\},  $r=\{0.5,1.5,\ldots,4.5$  fm\}, and $\theta=\{0, \pi/8, \ldots, \pi/2\}$. 
Here $A_i$ is the mass number of the $C_i$ cluster.
For $2\alpha+n(p)$ configurations,
$d=\{1.2,2.2,\ldots,6.2$ fm\}, $r=\{0.5,1.5,\ldots,6.5$ fm\}, and 
$\theta=\{0, \pi/8, \ldots, \pi/2\}$ are used to describe remarkable clustering  in $^9$Be($^9$B). 
For $\CO$, the $3\alpha+h$ configurations are restricted to be pyramid configurations with 
a regular triangle $3\alpha$ and a $h$ cluster on the vertical axis passing though the center of the 
$3\alpha$ plane. The side $d$ of the triangle and the hight $r$ are
chosen to be  $d=\{0.5,1.5,\ldots,3.5$ fm\} and $r=\{0.5,1.5,\ldots,4.5$ fm\}.

\subsubsection{AMD+VAP for $p$-shell nuclei}
The AMD+VAP method is applied for $p$-shell nuclei to 
investigate contributions of the spin-dependent $\La N$  interactions in $\LZ$. 
In the AMD framework, a basis wave function is given by a Slater determinant
\begin{equation}
 \Phi_{\rm AMD}({\bvec{Z}}) = \frac{1}{\sqrt{A_N!}} {\cal{A}} \{
  \varphi_1,\varphi_2,...,\varphi_{A_N} \},\label{eq:slater}
\end{equation}
where  ${\cal{A}}$ is the antisymmetrizer, and  $\varphi_i$ is 
the $i$th single-particle wave function written by a product of
spatial, spin, and isospin
wave functions, 
\begin{eqnarray}
 \varphi_i&=& \phi_{{\bvec{X}}_i}\chi_i\tau_i,\\
 \phi_{{\bvec{X}}_i}({\bvec{r}}_j) & = &  \left(\frac{2\nu}{\pi}\right)^{3/4}
\exp\bigl[-\nu({\bvec{r}}_j-\bvec{X}_i)^2\bigr],
\label{eq:spatial}\\
 \chi_i &=& (\frac{1}{2}+\xi_i)\chi_{\uparrow}
 + (\frac{1}{2}-\xi_i)\chi_{\downarrow},
\end{eqnarray}
where $\phi_{{\bvec{X}}_i}$ and $\chi_i$ are the spatial and intrinsic-spin functions, respectively, and 
$\tau_i$ is the isospin
function fixed to be proton or neutron. The width parameter $\nu$ is chosen to be the same value
as that used in the cluster model.  
The AMD wave function
is specified by a parameter set ${\bvec{Z}}\equiv 
\{{\bvec{X}}_1,\ldots, {\bvec{X}}_{A_N},\xi_1,\ldots,\xi_{A_N} \}$.
The Gaussian centroids ${\bvec{X}}_i$ and intrinsic-spin orientations  $\xi_i$ 
of all nucleons are independently treated as variational parameters in the 
energy variation. 
Owing to the  
flexibility of spatial and intrinsic-spin configurations of single-nucleon Gaussian wave packets, 
the AMD wave function can express various cluster structures with cluster breaking 
degrees of freedom as well as various 
intrinsic spin configurations.
Moreover, it can also describe shell-model wave functions because of the antisymmetrization.

In the AMD+VAP,  the energy variation is done after the  
angular-momentum and parity  projections in the AMD model space as  
\begin{eqnarray}
&& \frac{\delta}{\delta{\bvec{X}}_i}
\frac{\langle \Psi_N(I^\pi)|H_N|\Psi_N(I^\pi)\rangle}{\langle \Psi_N(I^\pi)|\Psi_N(I^\pi)\rangle}=0,\\
&& \frac{\delta}{\delta\xi_i}
\frac{\langle \Psi_N(I^\pi)|H_N|\Psi_N(I^\pi)\rangle}{\langle \Psi_N(I^\pi)|\Psi_N(I^\pi)\rangle}=0,\\
&&\Psi_N(I^\pi)= P^{I\pi}_{MK}\Phi_{\rm AMD}({\bvec{Z}}), 
\end{eqnarray}
in order to obtain the optimum solution of the parameter set  $\bvec{Z}$  for the lowest $I^\pi$ states.
For higher $I^\pi$ states, the VAP is done for the component orthogonal to the 
lower $I^\pi$ states already obtained by the VAP. The method is a version of the AMD and usually called the AMD+VAP. In this paper,  
it is simply denoted as the AMD. 

\subsubsection{Microscopic three-body model for $\CF$}
For $\CF$, we use the microscopic $^{16}\textrm{O}+p+n$  wave functions 
adopted in the previous study of $\CF$ \cite{Kanada-En'yo:2014oaa}. 
The wave functions are written in the form of the Brink-Bloch cluster wave functions 
for $C_1=p$, and $C_2=n$, and $C_3=^{16}\textrm{O}$ and are superposed with the GCM. 
The same parametrization of the generator coordinates as Ref.~\cite{Kanada-En'yo:2014oaa} is used,  
\begin{eqnarray}
\bvec{S}_1&=& (i q_x , r_y, 8D/9), \\
\bvec{S}_2&=& (-i q_x, -r_y, 8D/9),\\
\bvec{S}_3&=& (0, 0, -D/9).
\end{eqnarray}
In the present calculation, $D=\{1, 2, \ldots, 7$ fm\} are chosen. 
For all $D$ values, the coordinate set
$(q_x,r_y)=(0,0)$ is used corresponding to the $d$ cluster.
In addition, $q_x=\{0.5,1,1.5,2$ fm\} and  $r_y=\{0,1$ fm\} are used for 
$D=2$ fm and $q_x=\{1,2$ fm\} and  $r_y=\{0,1$ fm\} are used for 
$D=3,4,5$ fm to take into account the $d$-cluster breaking at the nuclear 
surface by the nuclear spin-orbit interactions from $^{16}\textrm{O}$.
In the total angular momentum projection $P^{J\pi}_{MK}$,  $|K|=1$ components are adopted. 
Higher $|K|$ states are not included 
to save computational costs in numerical integration of the Euler angles in the projection. 
The wave functions are automatically projected onto the isospin $T=0$ eigen states because of the $|K|=1$
projection. 
 
The inert $^{16}\textrm{O}$ cluster is assumed in the $^{16}\textrm{O}+p+n$  model.
In order to see possible 4$\alpha$-cluster vibration effect in $^{16}\textrm{O}$, a
$4\alpha+p+n$ model is also applied in the calculation of the leading $V_0$ contribution.
We use the label ``CL'' for the former model ($^{16}\textrm{O}+p+n$) 
with the inert $^{16}\textrm{O}$ cluster and 
CL-$4\alpha$ for the latter one ($4\alpha+p+n$) with the 
$^{16}\textrm{O}$ vibration.
In the CL-$4\alpha$ model,  regular tetrahedral $4\alpha$ configurations with 
the length of a side $r=0.5,1.5, 2.5$ fm are adopted in the GCM.
As shown later, the vibration effect is found to be  minor.

\subsubsection{Nuclear energy and density}
The Hamiltonian of the nuclear part consists of the kinetic terms, effective $NN$ interactions, and Coulomb interactions
as follows, 
\begin{equation}\label{eq:nuclear-H}
H_N=\sum^{A_N}_{i} \frac{1}{2m_N}\bvec{p}^2_i -T_G
+\sum^{A_N}_{i<j} V_{NN}(i,j)+\sum^Z_{i<j} V_\textrm{coulomb}(r_{ij})
\end{equation}
where $T_G$ is  the center of mass (cm)  kinetic energy, 
$V_{NN}$  is the effective $NN$ interactions, and $V_\textrm{coulomb}$ is the Coulomb interaction in the $A_N$-nucleon system.
The nuclear energy $E_N=\langle \Psi_N(I^\pi_n) |H_N|\Psi_N(I^\pi_n) \rangle$ 
and nuclear density  $\rho_N^{I^\pi}(r)$ are calculated for the 
nuclear wave functions $\Psi_N(I^\pi_n)$ (normalized as $|\langle \Psi_N(I^\pi_n) | \Psi_N(I^\pi_n) \rangle|=1$)
obtained with the CL and AMD models. 
In the calculation of the nuclear energy and density, the cm motion of core nuclei is removed exactly
and the radial coordinate $r$ in $\rho_N^{I^\pi}(r)$ is defined by the distance from the cm of core nuclei. 

\subsection{$\Lambda$ single-particle state with folding potential model in $\La$ hypernuclei}
\subsubsection{$\Lambda$ wave function}
The $\Lambda$ wave functions in $\La$ hypernuclei are calculated by 
assuming a $0s$-orbit $\La$ within a folding potential model
as done in the previous work \cite{Kanada-Enyo:2017ynk}. 
In the folding potential model, the $\La$-nucleus potential 
is obtained 
by folding the spin-independent $\La N$ central interactions 
($V_0$) with the nuclear density $\rho_N^{I^\pi}$.
The single-particle Hamiltonian $h_{\La,0}$ 
for a $0s$-orbit $\Lambda$ around the core is given as 
\begin{eqnarray}\label{eq:lambda-H}
h_{\La,0}&=&\frac{1}{2\mu_\Lambda}p_r^2+U_0(\rho_N^{I^\pi};r), \\
\mu_\Lambda&=&\frac{(A-1)m_N m_\Lambda}{(A-1)m_N + m_\Lambda}.
\end{eqnarray}
The nuclear density matrix in the exchange potential is approximated
with the density matrix expansion in the local density approximation  \cite{Negele:1975zz} as done in
the previous works. 
For a given nuclear density $\rho_N^{I^\pi}(r)$ of the core nucleus $I^\pi$ state, 
the single-particle energy $\varepsilon_{\La,0}(\rho_{N}^{I^\pi}
)$ and wave function 
$\phi_{\La,0}(\rho_N^{I^\pi};r)$ are 
calculated with the Gaussian expansion method \cite{Kamimura:1988zz,Hiyama:2003cu}.
In Appendix \ref{app:folding-pot}, we compare the approximately calculated $\La$-potential energies $\langle\phi_{\La,0} | U_0 (\rho)|\phi_{\La,0} \rangle$ 
in $^7\LLi(I^\pi=1^+,3^+)$ with those of the microscopically  calculated values 
 $\langle\Psi_N(I^\pi) \phi_{\La,0}  | V_0|\Psi_N(I^\pi) \phi_{\La,0} \rangle$.
It is found that the approximation in the exchange potential
gives only minor contribution to energy spectra.

In the cluster model calculation of $\LZ$ with $A\le 13$, 
we take into account the core polarization, i.e., the nuclear size change 
induced by the $\Lambda$ through the spin-independent $\La N$ central interactions ($V_0$)
in the same way as done in the previous works. 
It should be commented that, in the previous works, the core polarization  in $A>10$ hypernuclei was found to be small and gives 
only minor effect to energy spectra. The core polarization is omitted in the cluster model calculation of $\LO$ and $\LF$
for simplicity. 
In the AMD calculation, we adopt 
the frozen core approximation without the core polarization. 
Namely,  
we omit the core polarization and use the nuclear wave functions obtained for isolate $\CZ$ systems without the $\La$.

\subsection{Energy contributions of  spin-dependent $\La N$  interactions}

Energy contributions of the spin-dependent part $V_1$ of the $\La N$  interactions in $\LZ(J^\pi)$ 
are perturbatively calculated as $\langle \Psi_{\LZ}(J^\pi) | V_1 | \Psi_{\LZ}(J^\pi)\rangle$
by using  the unperturbative $A$-body 
microscopic wave functions
\begin{eqnarray}
\Psi_{\LZ}(J^\pi)= \left[\Psi_N(I^\pi_n)  \phi_{\La,0}\chi_\La \right ]_J,
\end{eqnarray}
obtained with $V_0$.
Here $\chi_\La$ is the  $\La$-spin function coupling with 
the nuclear spin $I$ to the total 
angular momentum $J$. 
In the present perturbative treatment,   modification of the 
$\La$ wave function by $V_1$ is omitted.
Moreover,  rearrangement of spin configurations by the $V_1$ contribution is ignored. 
We checked the spin rearrangement effect to energy spectra 
and found that it is mostly minor unless 
energies of two $J^\pi$ states for different $I^\pi$ are 
close to each other.


\subsection{Calculation procedure}

The procedures of the present calculation are summarized below.
The calculation is  done in three steps as follows.
\begin{enumerate}
\item Calculation of the core nuclear part is performed with two kinds of structure models, 
cluster and AMD models. (For $\LF$, two kinds of cluster model calculations are done.) 
The nuclear wave function $\Psi_N(I^\pi)$, nuclear energy $E_N(I^\pi)$, and nuclear density
$\rho_N^{I^\pi}(r)$  for $\CZ(I^\pi)$ are obtained with the Hamiltonian given by 
Eq.~\eqref{eq:nuclear-H}. 

\item Using the nuclear density $\rho_N^{I^\pi}$ obtained in the first step, 
the $0s$-orbit $\Lambda$ state in  $\LZ$ is calculated by the folding potential model 
with the spin-independent part  $V_0$ of the  $\La N$ interactions.
The  $\La$ single-particle energy 
$\varepsilon_{\La,0}(\rho_{N}^{I^\pi})$ and the $\La$ wave function
$\phi_{\La,0}(\rho_N^{I^\pi})$ 
are obtained with the Hamiltonian given by Eq.~\eqref{eq:lambda-H}.
The core polarization is taken into account in the cluster model calculations of
$\LZ$ with $A\le 13$. It is omitted in other calculations.
\item 
The energy contributions from the spin-dependent part $V_1$ of the $\La N$ interactions  are 
perturbatively calculated with the unperturbative wave functions
$\Psi_{\LZ}(J^\pi)$ given by $\Psi_N(I^\pi_n)$ and 
$\phi_{\La,0}$, which are obtained in the first and second steps, respectively. 
The spin rearrangement and $\phi_{\La,0}$ change induced by $V_1$ are ignored. 
The third step calculation is done by using the nuclear wave functions $\Psi_N(I^\pi_n)$ obtained with 
the AMD (for $\LF$, that with the cluster model).
\end{enumerate}

\subsection{Energies}
The $\La N$ interactions ($V_{\La N}$) contains the spin-independent part ($V_0$) and the
spin-dependent part ($V_1$). 
We take into account four spin-dependent terms of $V_1$  as
\begin{eqnarray}
V_1&=&V_{\sigma} + V_{S_\La}+V_{S_N}+
V_T,
\end{eqnarray}
where $V_{\sigma}$, $V_{S_\La}$,  $V_{S_N}$, and $V_T$ are 
the spin-spin  ($\bvec{\sigma}_\La\cdot \bvec{\sigma}_N$), the $\La$-spin spin-orbit ($\bvec{l}\cdot\bvec{s}_\La$), 
the nucleon-spin spin-orbit ($\bvec{l}\cdot\bvec{s}_N$), and the tensor ($ S_{12}$) terms, respectively. 
Following the notation usually used in the shell model (SM) calculations \cite{Millener:2008zz,Millener:2010zz,Millener:2012zz}, 
we denote the contributions from $V_{\sigma}$, $V_{S_\La}$, $V_{S_N}$, and $V_T$ terms as  
the $\Delta_\sigma$, $S_\La$, $S_N$, and $T$ contributions, respectively. 
The $\La\Sigma$ contribution 
from the  $\La\Sigma$ coupling is ignored in the present calculation.
In this section, we explain definitions of energies and respective contributions.

\subsubsection{Total energy and excitation energy}
The total energy of $\LZ$ is given as 
\begin{equation}
E_{\LZ}(J^\pi)=E_N(I^\pi)+\varepsilon_{\La,0}(\rho_N^{I^\pi})
+\langle\Psi_{\LZ}(J^\pi)|V_1|\Psi_{\LZ}(J^\pi)\rangle.
\end{equation}
We regard the sum of the first and second terms as the leading term contributed from
$V_0$ and the third term as the perturbative term from $V_1$. 
In the present paper, 
we evaluate  energies of $\LZ$ in two ways as follows. 
In the first calculation, we evaluate both the leading and the perturbative terms 
with the AMD model calculation as  
\begin{eqnarray}
E_{\LZ}(J^\pi)&=&E^\textrm{AMD}_N(I^\pi)+\varepsilon^\textrm{AMD}_{\La,0}(\rho_N^{I^\pi})\nonumber\\
&+&\langle\Psi^\textrm{AMD}_{\LZ}(J^\pi)|V_1|\Psi^\textrm{AMD}_{\LZ}(J^\pi)\rangle,
\end{eqnarray}
where $E^\textrm{AMD}_N$, $\varepsilon^\textrm{AMD}_{\La,0}$, and $\Psi^\textrm{AMD}_{\LZ}$
are the nuclear energy, $\La$ single-particle energy, and $\LZ$ wave function obtained by the AMD.
This is a consistent calculation, in which the $V_0$ and $V_1$ contributions are calculated with the AMD.
However, inter-cluster motion is not necessarily described sufficiently in the present AMD calculation
because of the number of basis wave functions is limited as mentioned previously. 
The inter-cluster motion may give important contributions, in particular, to the leading term.
Therefore, we also adopt an alternative way of energy evaluation 
by replacing the leading terms with those obtained by the CL calculation as 
\begin{eqnarray}
E_{\LZ}(J^\pi)&=&E^\textrm{CL}_N(I^\pi)+\varepsilon^\textrm{CL}_{\La,0}(\rho_N^{I^\pi})\nonumber\\
&+&\langle\Psi^\textrm{AMD}_{\LZ}(J^\pi)|V_1|\Psi^\textrm{AMD}_{\LZ}(J^\pi)\rangle,
\end{eqnarray}
which we  call the  ``CL+AMD'' calculation. Here $E^\textrm{CL}_N$ and $\varepsilon^\textrm{CL}_{\La,0}$
are the nuclear energy and $\La$ single-particle energy obtained by the CL with $V_0$.

In both the AMD and CL+AMD calculations, excitation energies are given 
by the energy difference between the ground and excited states as
\begin{equation}
E_x(J^\pi)=E_{\LZ}(J^\pi)-E_{\LZ}(\textrm{gs}).
\end{equation}

\subsubsection{The spin-averaged energies}
The spin-averaged energy $\bar{E}_{\LZ}(I^\pi)$ is defined by the 
energy averaged for the spin-doublet partners  $J^\pi_>$ and $J^\pi_<$  in $\LZ$ states for the core . $\CZ(I^\pi)$ state.
It is given by sum of the leading and perturbative terms 
corresponding to the $V_0$ and $V_1$ contributions as 
\begin{eqnarray}
\bar{E}_{\LZ}(I^\pi)&=&\bar{E}_{\LZ,0}(I^\pi)+\langle V_{S_N} \rangle_{\LZ(I^\pi)},\\
\bar{E}_{\LZ,0}(I^\pi)&=&E_N(I^\pi)+\varepsilon_{\La,0}(\rho_N^{I^\pi}),\\
\langle V_{S_N} \rangle_{\LZ(I^\pi)}&=&
\langle\Psi_{\LZ}(J^\pi)|V_{S_N}|\Psi_{\LZ}(J^\pi)\rangle.
\end{eqnarray}
The perturbative term is contributed by the 
$S_N$ term of the spin-dependent $\La N$ interactions
($V_{S_N}$ in $V_1$). 
Note that, in the present calculation without the spin rearrangement, 
the $S_N$ contribution has no $\La$-spin dependence and
depends only on  the core nuclear spin $I^\pi$.
The spin-averaged binding energy $\bar{B}_{\LZ}(I^\pi)$ 
is given by the sum of 
leading and perturbative terms as 
\begin{eqnarray}
\bar{B}_\La(I^\pi)&=&\bar{B}_{\La,0}(I^\pi)+\bar{B}_{\La,S_N}(I^\pi),\\
\bar{B}_{\La,0}(I^\pi)&=&E_{\CZ}(I^\pi)- \bar{E}_{\LZ,0}(I^\pi),\\
\bar{B}_{\La,S_N}(I^\pi)&=&-\langle V_{S_N} \rangle_{\LZ(I^\pi)},
\end{eqnarray}
where $\bar{B}_{\La,0}$ and $\bar{B}_{\La,S_N}$ are the $V_0$ and $S_N$ contributions
to $\bar{B}_\La$, respectively.
The spin-averaged excitation energy $\bar{E}_x(I^\pi)$ is defined as 
\begin{equation}
\bar{E}_x(I^\pi)=\bar{E}_{\LZ}(I^\pi)-\bar{E}_{\LZ}(\textrm{gs}).
\end{equation}
We define the spin-averaged excitation energy shift $\delta_\La(\bar{E}_x)$ induced by the $\La$ 
in $\LZ$ by the difference from the original excitation energy  $E_x(\CZ)$ in the ordinary nucleus 
$\CZ$ as 
\begin{equation}
\delta_\La (\bar{E}_x(I^\pi))=\bar{E}_x(I^\pi)-E_x(\CZ;I^\pi).
\end{equation}
The energy shift $\delta_\La(\bar{E}_x)$ can be separated into two components 
of the $V_0$ and $S_N$ contributions 
\begin{equation}
\delta_\La (\bar{E}_x(I^\pi))=\delta_{\La,0} (\bar{E}_x(I^\pi))+\delta_{\La,S_N}(\bar{E}_x(I^\pi)).\label{eq:ex-shift}
\end{equation}
In the calculation without the core polarization, the first  and second components are given by energy differences in
$\varepsilon_{\La,0}$ and  $S_N$ contributions, respectively,  between the ground and excited states
as 
\begin{eqnarray} \label{eq:energy-shift-v0-sn}
\delta_{\La,0}(\bar{E}_x)&=& \varepsilon_{\La,0}(\rho_N^{I^\pi}) -
\varepsilon_{\La,0}(\rho_N^\textrm{gs}), \\
\delta_{\La,S_N}(\bar{E}_x)&=&
\langle V_{S_N} \rangle_{\LZ(I^\pi)}
\nonumber\\
&&-
\langle V_{S_N} \rangle_{\LZ(\textrm{gs})}.
\end{eqnarray}

\subsubsection{spin-doublet splitting energy}
The splitting energy 
between spin-doublet  $J^\pi_>$ and  $J^\pi_<$ states 
is given as 
\begin{equation}
E_{\LZ}(J^\pi_>)-E_{\LZ}(J^\pi_<).
\end{equation}
Here the splitting energy is defined by the energy  of the $J^\pi_>$ state 
measured from the $J^\pi_<$ state, and a negative splitting energy means the 
reverse ordering case $E_{\LZ}(J^\pi_>)<  E_{\LZ}(J^\pi_<)$.
In the present perturbative treatment, 
the splitting energy is sum of  $\Delta_\sigma$, $S_\La$, and $T$ contributions,
\begin{eqnarray}
\langle  \Psi_{\LZ}(J^\pi_>)| V_{\sigma,S_\La,T} |\Psi_{\LZ}(J^\pi_>) \rangle
-\langle \Psi_{\LZ}(J^\pi_<)| V_{\sigma,S_\La,T}|\Psi_{\LZ}(J^\pi_<)\rangle. \nonumber
\end{eqnarray}

 
\section{Effective  interactions}\label{sec:interactions}

\subsection{Effective $NN$ interactions}
As for the effective $NN$ interactions, 
the finite-range $NN$ interactions of 
the Volkov central   and G3RS spin-orbit 
 forces  \cite{VOLKOV,LS} are adopted. These interactions are widely used  in structure studies of $p$-shell nuclei.

\begin{table}[thb]
\caption{
Parameter sets of the Volkov+G3RS	$NN$ interactions and 
the $\bar{k}_F$ parameters of the ESC08a $\La NG$ interaction
in the CL, AMD, and AMD' calculations with the hybrid $k_F$ treatment. 
\label{tab:input-parameters}
}
\begin{center}
\begin{tabular}{c|cc|cc|ccccccccc}
& 		\multicolumn{3}{c}{$NN$ central} &	\multicolumn{3}{c}{$NN$ spin-orbit}\\
set & 		\multicolumn{3}{c}{Volkov No.2} &	\multicolumn{3}{c}{G3RS [MeV]}\\
& \\
NN-a	&\multicolumn{3}{c}{$m=0.60$, $b=h=0.125$}	&	\multicolumn{3}{c}{$u_1=-u_2=1600$} \\
NN-a'	&\multicolumn{3}{c}{$m=0.60$, $b=h=0.06$}	 &		\multicolumn{3}{c}{$u_1=-u_2=1300$} \\
NN-b	&\multicolumn{3}{c}{$m=0.57$, $b=h=0.125$}	&		\multicolumn{3}{c}{$u_1=-u_2=1200$} \\
NN-c	&\multicolumn{3}{c}{$m=0.62$, $b=h=0.125$}	&		\multicolumn{3}{c}{$u_1=-u_2=820$} \\
& \\
\hline
        &\multicolumn{2}{c|}{ CL}
&\multicolumn{2}{c|}{AMD}&\multicolumn{2}{c}{ AMD'} \\
 $\LZ$		&	\qquad 	set\qquad &	\quad $\bar{k}_F$	\quad &\qquad 	set	\qquad&	\quad $\bar{k}_F$	\quad&\qquad 	set	\qquad &	\quad $\bar{k}_F$\quad	\\
$^7\LLi$	&		NN-b	&	0.93 	&	NN-b	&	0.96 	&		&		\\
$^8\LLi$	&		NN-b	&	0.91 	&	NN-b	&	0.96 	&		&		\\
$^9\LBe$	&		NN-a	&	0.90 	&	NN-a	&	0.96 	&	NN-a'	&	0.95 	\\
$^{10}\LBe$	&	NN-a	&	0.95 	&	NN-a	&	0.97 	&	NN-a'	&	0.98 	\\
$^{10}\LB$	&	NN-a	&	0.94 	&	NN-a	&	0.96 	&	NN-a'	&	0.97 	\\
$^{11}\LBe$	&	NN-a	&	1.04 	&	NN-a	&	1.10 	&	NN-a'	&	1.05 	\\
$^{11}\LB$	&		NN-a	&	1.03 	&	NN-a	&	1.06 	&	NN-a'	&	1.06 	\\
$^{12}\LB$	&		NN-a	&	1.07 	&	NN-a	&	1.12 	&	NN-a'	&	1.11 	\\
$^{12}\LC$	&		NN-a	&	1.06 	&	NN-a	&	1.12 	&	NN-a'	&	1.10 	\\
$^{13}\LC$	&		NN-a	&	1.11 	&	NN-a	&	1.18 	&	NN-a'	&	1.15 	\\
$\LO$	&		NN-a	&	1.14 	&	NN-a	&	1.21 	&	NN-c	&	1.20 	\\
$\LF$	&		NN-c	&	1.18 	&		&		&		&		\\
\hline
\end{tabular}
\end{center}
\end{table}

These effective $NN$ interactions have adjustable parameters, which are usually tuned $A_N$- and model-dependently. 
The interaction parameters used in the present calculation are summarized in Table \ref{tab:input-parameters}.
The default parameter sets are the same as those used  for the cluster model calculations in the previous works
\cite{Kanada-Enyo:2017ynk,Kanada-Enyo:2018pxt}.
They  are $w=0.40$, $m=0.60$, and $b=h=0.125$ of  the Volkov No.2 and 
$u_1=-u_2=1600$ MeV of the G3RS for Be, B, C, and O, and 
$w=0.43$, $m=0.57$, $b=h=0.125$, and $u_1=-u_2=1200$ MeV for $^6$Li and $^7$Li.
In this paper, the former and the latter sets are called NN-a and NN-b, respectively. 
For $\LF$, the set (labeled by NN-c) of $w=0.38$, $m=0.62$, $b=h=0.125$, and $u_1=-u_2=820$ MeV is used.  
The NN-c was tuned so as to describe spectra in $^{17}\textrm{O}$ and $\CF$ with the $^{16}\textrm{O}+n$ and 
 $^{16}\textrm{O}+p+n$ cluster model calculations \cite{Kanada-En'yo:2014oaa}. 
The CL calculation with these default parameters of the $NN$ interactions globally describes low-lying energy spectra in core nuclei. 
However, the AMD calculation with the default parameters sometimes fails to reproduce 
the experimental energy spectra in such nuclei as $^{10}\CB$ and $\CO$. Instead of the NN-a, 
we also use an alternative set (NN-a') of modified parameters  $w=0.40$, $m=0.60$, $b=h=0.06$, and 
$u_1=-u_2=1300$ MeV, which have been used in the AMD calculation of $^{10}\CB$ \cite{Kanada-Enyo:2015uiy}.
For the AMD calculation of $\CO$, we try the NN-c. 
In most cases, the modifications of $NN$ interaction parameters give only minor 
changes of nuclear structures except for energy spectra. 
We use the label AMD for the AMD calculation with the default parameters of the $NN$ interactions, 
and the label AMD' for the AMD calculation with the set NN-a' 
(for $\LO$, the label AMD' for the calculation 
with the set NN-c). 

The parameter sets of effective $NN$ interactions are summarized in Table \ref{tab:input-parameters}.

\subsection{$\Lambda$-$N$ interactions}
We use the $\La N G$ interactions, which are derived  with the $G$-matrix theory from the 
the Nijmegen extended-soft-core (ESC) model. 
The $\La NG$ interactions used here are 
even and odd central, triplet-odd spin-orbit, and triplet-odd tensor interactions  as
\begin{eqnarray}
V_{\La N}&=&V_0+V_{\sigma} + V_{S_\La}+V_{S_N}+V_T,\\
V_0&=&v^\textrm{e}_0(r) \hat P(E) 
+ v^\textrm{o}_0(r) \hat P(O) \\
V_{\sigma} &=&\left [v^\textrm{e}_{\sigma}(r) \hat P(E)
+ v^\textrm{o}_{\sigma}(r) \hat P(O) \right](\bvec{\sigma}_\La\cdot \bvec{\sigma}_N)\\
V_{S_\La} &=& 
v^\textrm{o}_{S_\La}(r)\hat P(O)  (\bvec{l}\cdot\bvec{s}_\Lambda )
\\
V_{S_N} &=& 
v^\textrm{o}_{S_N}(r)\hat P(O)  (\bvec{l}\cdot\bvec{s}_N)
\\
V_T&=&v^\textrm{o}_T(r)  \hat P(O) S_{12}
\\
\hat P(E)&=&\frac{1+P_r}{2} \\
\hat P(O)&=& \frac{1-P_r}{2}.
\end{eqnarray}
For simplicity, 
the $\La$-$N$ relative momentum $\bvec{p}$ 
in the $\bvec{l}=\bvec{r}\times \bvec{p}$ term of the 
spin-orbit interactions $V_{S_\La}$ and  $V_{S_N}$ is approximated to be 
$\bvec{p}=(\bvec{p}_N-\bvec{p}_\La)/2$ corresponding to the equal mass ($m_N=m_\La$) approximation.

We start from the ESC08a version of the $\Lambda NG$ interactions 
\cite{Rijken:2010zza,Rijken:2010zzb,Yamamoto:2010zzn}, and then consider tuning of
its spin-dependent terms as described in later. 
In Ref.~\cite{Yamamoto:2010zzn}, 
the original ESC08a  $\Lambda NG$ interaction is given by  three-range Gaussian
local potentials.
$v^\textrm{e,o}_{0,\sigma}(r)$ for the central interactions $V_0$ and 
$V_\sigma$ are written as 
\begin{eqnarray}
&&v^\textrm{e,o}_{0,\sigma} (k_F, r)= \sum_{i=1}^3 \sum_{n=0}^2
c^\textrm{e,o}_{0,\sigma} (n,i)  k_F ^n
\exp\left[-\left(\frac{r}{\beta_i}\right)^2\right], \\
&&c^\textrm{e}_0(n,i)=c^\textrm{1E}_{n,i}+c^\textrm{3E}_{n,i},\\
&&c^\textrm{o}_0(n,i)=c^\textrm{1O}_{n,i}+c^\textrm{3O}_{n,i},\\
&&c^\textrm{e}_\sigma(n,i)=c^\textrm{3E}_{n,i}-3c^\textrm{1E}_{n,i},\\
&&c^\textrm{o}_\sigma(n,i)=c^\textrm{3O}_{n,i}-3c^\textrm{1O}_{n,i},
\end{eqnarray}
with the Gaussian range parameters $\beta_1=0.5$ fm, $\beta_2=0.9$ fm,  and $\beta_3=2.0$ fm. 
The density dependence is taken into account by the $k_F$ parameter. 
The values of $c^\textrm{1E}_{n,i}$, $c^\textrm{3E}_{n,i}$, $c^\textrm{1O}_{n,i}$,
and $c^\textrm{3E}_{n,i}$ for the ESC08a are listed in Table II of Ref.~\cite{Yamamoto:2010zzn}.

\begin{table}[ht]
\caption{Parameters of triplet-odd $\La$-spin spin-orbit ($V_{S_\La}$) 
and tensor ($V_T$) terms 
in the ESC08a $\La NG$ interaction at $k_F=1.0$ fm$^{-1}$ from 
Ref.~\cite{Yamamoto:2010zzn}.
\label{tab:LS-T} }
\begin{center}
\begin{tabular}{ccccccc}
\hline
& $i=1$ & $i=2$ & $i=3$ \\
\multicolumn{4}{l}{$V_{S_\La}$}\\
$\beta'_i$	[fm] &$	0.4	$&$	0.8	$&$	1.2	$\\
$c^\textrm{o}_{S_\La}$ [MeV]  &$	2772	$&$	-106.6	$&$	-0.864	$\\
$c^\textrm{o}_{S_N}$  [MeV] &$	-684	$&$	-129.8	$&$	-4.538	$\\
&\\
\multicolumn{4}{l}{$V_{T}$}\\
$\beta_i$[fm]	&$	0.5	$&$	0.9	$&$	2	$\\
$c^\textrm{o}_{T}$ [MeV] &$	10.37	$&$	0.0181	$&$	0.017$	\\
\hline		
\end{tabular}
\end{center}
\end{table}

For the spin-orbit and tensor interactions, we use 
the density-independent interactions fixed at $k_F=1.0$ fm$^{-1}$ as
\begin{eqnarray}
v^\textrm{o}_{S_\La,S_N}(r)&=&  \sum_{i=1}^3 
c^\textrm{o}_{S_\La,S_N}(i) \exp\left[-\left(\frac{r}{\beta'_i}\right)^2\right], \\
v^\textrm{o}_{T}(r)&=&\sum_{i=1}^3 
c^\textrm{o}_{T}(i) r^2\exp\left[-\left(\frac{r}{\beta_i}\right)^2\right].
\end{eqnarray}
The values of range and strength parameters taken from Ref.~\cite{Yamamoto:2010zzn}
are listed in Table \ref{tab:LS-T}.

For the $k_F$ parameter in the central interactions, 
there are a couple of treatments. 
One is the  
density-dependent (DD) $k_F$ treatment called ``averaged density approximation (ADA)'', 
and another is the density-independent (DI) $k_F$ treatment with a fixed $k_F$ value.
The $\La NG$ interactions generally have density dependence reflecting nuclear medium effects, 
which are taken into account in the $G$-matrix theory.
ESC08 versions of the $\La NG$ interactions was originally designed as density-dependent 
interactions to globally reproduce the $\La$ binding energies of $\LZ$ 
in a wide mass number region ~\cite{Yamamoto:2010zzn, Isaka:2016apm,Isaka:2017nuc}, 
whereas the DI  $k_F$ treatment has been often used in studies of 
energy spectra of $p$-shell hypernuclei.
In the previous works \cite{Kanada-Enyo:2017ynk,Kanada-Enyo:2018pxt}, 
applicability of the DD and DI $k_F$ treatments for description of $p$-shell $\LZ$ energy spectra
has been tested focusing on excitation energy shifts by $\La$ particle.
The DD $k_F$ treatment is found to be not suitable to describe the observed excitation energy shifts in $\LZ$.
The DI $k_F$ treatment can describe a trend of the excitation energy shifts  but tends to overestimate the 
observed values. It suggests that a moderate density-dependence weaker than the DD $k_F$ treatment 
is favored, and therefore the
intermediate version (hybrid $k_F$ treatment) 
between DD and DI 
has been proposed as an alternative $k_F$ treatment in Ref.~\cite{Kanada-Enyo:2018pxt}. 
In the present calculation, we adopt the hybrid $k_F$ treatment described as follows.

In the DD $k_F$ treatment (ADA), 
the $k_F$ is taken to be
$k_F=\langle k_F \rangle_\Lambda$, where 
$\langle k_F \rangle_\Lambda$ is the averaged Fermi momentum for the $\Lambda$ distribution as 
\begin{equation}
\langle k_F \rangle_\Lambda
=\left[\frac {3\pi^2}{2}\langle \rho_N \rangle_\Lambda \right]^{1/3}.
\end{equation}
In the hybrid $k_F$ treatment, the average of the  DD and DI  interactions are used. 
Namely, the $k_F$ parameter is chosen to be 
\begin{eqnarray}
k^n_F= (1-e) \bar{k}^n_F+e\langle k_F\rangle_\La^n, 
\end{eqnarray}
with a weight factor $e=0.5$. Here $\bar{k}_F$ is the fixed input parameter.
In the DD and hybrid $k_F$ treatments, 
$\langle k_F \rangle_\Lambda$ is self-consistently 
determined for each state in the $\La$-nucleus potential model. In the hybrid $k_F$ treatment, 
the input parameter $\bar{k}_F$ is chosen for each system and taken to be 
the mean value of $\langle k_F \rangle_\Lambda$ determined by the DD $k_F$ treatment for the
ground and excited states of the system. 
The used values of the input parameter $\bar{k}_F$ in the hybrid $k_F$ treatment 
are shown in Table \ref{tab:input-parameters}.


\section{Properties of core nuclei $\CZ$}  \label{sec:CZ}

In order to investigate energy spectra in $\LZ$, 
it is important that the structure models properly reproduce the
structure properties such as energy spectra and radii of core nuclear states ($\CZ$) without a $\La$.
In particular, nuclear spin properties of core nuclei $\CZ$ are essential to discuss 
spin-dependent contributions of the $\La N$ interactions 
in $\LZ$. In this section, we show the calculated result for $\CZ$. 
In addition to  energy spectra of $\CZ$, 
a particular attention is paid on spin configurations, which are directly reflected in 
spin-doublet splittings in $\LZ$.

\subsection{Energies and radii of core nuclei $\CZ$}

\begin{table*}[ht]
\caption{Nuclear properties of excitation energies ($E_x$ [MeV]), magnetic moments ($\mu$ [$\mu_N$]), and intrinsic-spin and orbital 
angular momentum expectation values 
in $\CZ$. The AMD and AMD' results are shown. For excitation energies, The CL result of $E_x$ is also shown. 
For $\CF$, the CL result  is shown. The experimental data are from 
Refs.~\cite{AjzenbergSelove:1990zh,Tilley:2002vg,Tilley:2004zz,Kelley:2012qua} 
\label{tab:mu}
}
\begin{center}
\begin{tabular}{cc|cccc|c|cccc|ccccccc}
\hline			
$\CZ$ &  $I^\pi$ 	 &exp & CL& AMD & AMD' &exp &\multicolumn{4}{c|}{AMD} &\multicolumn{4}{c}{AMD'}  \\
& &$E_x$ &  $E_x$ & $E_x$ & $E_x$ 
& $	\mu$&$\mu	$&$\langle S_z\rangle	$&$	\langle S^2\rangle	$&$		\langle L^2\rangle	$&$\mu	$
&$\langle S_z\rangle	$&$	\langle S^2\rangle	$&$		\langle L^2\rangle	$\\
$^6\CLi$	&$	3^+	$&$	2.186 	$&$	2.08 	$&$	2.00 	$&$		$&$	-	$&$	1.88 	$&$	1.00 	$&$	2.00 	$&$	6.00 	$&$		$&$		$&$		$&$		$\\
$^6\CLi$	&$	1^+_\textrm{gs}	$&$		$&$		$&$		$&$		$&$	0.822 	$&$	0.88 	$&$	1.00 	$&$	2.00 	$&$	0.00 	$&$		$&$		$&$		$&$		$\\
&&&&&&&&&&&&&& \\
$^7\CLi$	&$	7/2^-	$&$	4.630 	$&$	4.75 	$&$	4.88 	$&$		$&$	-	$&$	3.74 	$&$	0.52 	$&$	0.82 	$&$	11.92 	$&$		$&$		$&$		$&$		$\\
$^7\CLi$	&$	5/2^-	$&$	6.680 	$&unbound	&$	7.17 	$&$		$&$	-	$&$	-0.99 	$&$	-0.35 	$&$	0.77 	$&$	11.96 	$&$		$&$		$&$		$&$		$\\
$^7\CLi$	&$	3/2^-_\textrm{gs}	$&$		$&$		$&$		$&$		$&$	3.256 	$&$	3.13 	$&$	0.50 	$&$	0.79 	$&$	2.04 	$&$		$&$		$&$		$&$		$\\
$^7\CLi$	&$	1/2^-	$&$	0.478 	$&$	0.49 	$&$	0.79 	$&$		$&$	-	$&$	-0.74 	$&$	-0.16 	$&$	0.78 	$&$	2.02 	$&$		$&$		$&$		$&$		$\\
&&&&&&&&&&&&&& \\
$^8\CBe$	&$	2^+	$&$	3.040 	$&$	3.11 	$&$	3.34 	$&$	3.34 	$&$	-	$&$	1.00 	$&$	0.00 	$&$	0.02 	$&$	6.01 	$&$	1.00 	$&$	0.00 	$&$	0.01 	$&$	6.01 	$\\
$^8\CBe$	&$	0^+_\textrm{gs}$&$		$&$		$&$		$&$		$&$	-	$&$	0.00 	$&$	0.00 	$&$	0.02 	$&$	0.02 	$&$	0.00 	$&$	0.00 	$&$	0.02 	$&$	0.02 	$\\
&&&&&&&&&&&&&& \\
$^9\CBe$	&$	5/2^-	$&$	2.429 	$&$	2.02 	$&$	2.27 	$&$	2.22 	$&$	-	$&$	-0.78 	$&$	0.44 	$&$	0.76 	$&$	6.46 	$&$	-0.89 	$&$	0.46 	$&$	0.76 	$&$	6.29 	$\\
$^9\CBe$	&$	3/2^-_\textrm{gs}	$&$		$&$		$&$		$&$		$&$	-1.178 	$&$	-1.13 	$&$	0.37 	$&$	0.76 	$&$	2.64 	$&$	-1.23 	$&$	0.40 	$&$	0.76 	$&$	2.52 	$\\
$^9\CBe$	&$	1/2^-	$&$	2.780 	$&$	2.20 	$&$	3.23 	$&$	2.53 	$&$	-	$&$	0.83 	$&$	-0.17 	$&$	0.76 	$&$	2.00 	$&$	0.84 	$&$	-0.17 	$&$	0.75 	$&$	2.00 	$\\
&&&&&&&&&&&&&& \\
$^{10}\CBe$	&$	2^+	$&$	3.368 	$&$	3.21 	$&$	3.67 	$&$	3.51 	$&$	-	$&$	1.11 	$&$	0.11 	$&$	0.75 	$&$	6.09 	$&$	0.83 	$&$	0.05 	$&$	0.32 	$&$	6.03 	$\\
$^{10}\CBe$	&$	0^+_\textrm{gs}	$&$		$&$		$&$		$&$		$&$	-	$&$	0.00 	$&$	0.00 	$&$	0.74 	$&$	0.74 	$&$	0.00 	$&$	0.00 	$&$	0.37 	$&$	0.37 	$\\
&&&&&&&&&&&&&& \\
$^{10}\CB$	&$	3^+_\textrm{gs}	$&$		$&$		$&$		$&$		$&$	1.801 	$&$	1.82 	$&$	0.85 	$&$	2.07 	$&$	7.25 	$&$	1.83 	$&$	0.86 	$&$	2.03 	$&$	7.14 	$\\
$^{10}\CB$	&$	1^+	$&$	0.718 	$&$	1.21 	$&$	4.15 	$&$	1.87 	$&$	0.63(2)	$&$	0.76 	$&$	0.69 	$&$	2.01 	$&$	1.24 	$&$	0.77 	$&$	0.70 	$&$	2.01 	$&$	1.21 	$\\
&&&&&&&&&&&&&& \\
$^{11}\CB$	&$	5/2^-	$&$	4.445 	$&$	4.66 	$&$	3.98 	$&$	2.87 	$&$	-	$&$	3.77 	$&$	0.51 	$&$	1.15 	$&$	6.37 	$&$	3.81 	$&$	0.49 	$&$	0.91 	$&$	6.24 	$\\
$^{11}\CB$	&$	3/2^-_\textrm{gs}	$&$		$&$		$&$		$&$		$&$	2.689 	$&$	2.32 	$&$	0.26 	$&$	1.26 	$&$	3.74 	$&$	2.27 	$&$	0.25 	$&$	0.95 	$&$	3.43 	$\\
$^{11}\CB$	&$	1/2^-	$&$	2.125 	$&$	2.79 	$&$	2.77 	$&$	1.22 	$&$	-	$&$	-0.61 	$&$	-0.16 	$&$	0.98 	$&$	2.22 	$&$	-0.60 	$&$	-0.16 	$&$	0.87 	$&$	2.11 	$\\
$^{11}\CB$	&$	3/2^-_2	$&$	5.020 	$&$	5.57 	$&$	6.09 	$&$	4.08 	$&$	-	$&$	0.75 	$&$	-0.01 	$&$	1.00 	$&$	4.75 	$&$	0.69 	$&$	-0.04 	$&$	0.85 	$&$	4.79 	$\\
&&&&&&&&&&&&&& \\
$^{11}\CC$	&$	5/2^-	$&$	4.319 	$&$	4.50 	$&$	3.87 	$&$	2.84 	$&$	-	$&$	-0.88 	$&$	0.50 	$&$	1.14 	$&$	6.38 	$&$	-0.93 	$&$	0.49 	$&$	0.90 	$&$	6.24 	$\\
$^{11}\CC$	&$	3/2^-_\textrm{gs}	$&$		$&$		$&$		$&$		$&$	-0.964 	$&$	-0.63 	$&$	0.26 	$&$	1.25 	$&$	3.74 	$&$	-0.58 	$&$	0.26 	$&$	0.94 	$&$	3.41 	$\\
$^{11}\CC$	&$	1/2^-	$&$	2.000 	$&$	2.62 	$&$	2.64 	$&$	1.16 	$&$	-	$&$	0.99 	$&$	-0.16 	$&$	0.97 	$&$	2.21 	$&$	0.98 	$&$	-0.16 	$&$	0.86 	$&$	2.10 	$\\
$^{11}\CC$	&$	3/2^-_2	$&$	4.804 	$&$	5.35 	$&$	5.91 	$&$	4.01 	$&$	-	$&$	0.76 	$&$	-0.01 	$&$	0.99 	$&$	4.75 	$&$	0.80 	$&$	-0.04 	$&$	0.84 	$&$	4.79 	$\\
&&&&&&&&&&&&&& \\
$^{12}\CC$	&$	2^+	$&$	4.439 	$&$	4.47 	$&$	4.70 	$&$	6.07 	$&$	-	$&$	1.03 	$&$	0.06 	$&$	0.50 	$&$	6.14 	$&$	1.00 	$&$	0.03 	$&$	0.22 	$&$	6.05 	$\\
$^{12}\CC$	&$	0^+_\textrm{gs}	$&$		$&$		$&$		$&$		$&$	-	$&$	0.00 	$&$	0.00 	$&$	0.95 	$&$	0.95 	$&$	0.00 	$&$	0.00 	$&$	0.28 	$&$	0.28 	$\\
&&&&&&&&&&&&&& \\
$\CO$	&$	3/2^-	$&$	6.176 	$&$	5.65 	$&$	9.88 	$&$	0.96 	$&$	-	$&$	-1.82 	$&$	0.49 	$&$	0.81 	$&$	2.11 	$&$	-1.83 	$&$	0.49 	$&$	0.79 	$&$	2.09 	$\\
$\CO$	&$	1/2^-_\textrm{gs}	$&$		$&$		$&$		$&$		$&$	0.719 	$&$	0.64 	$&$	-0.16 	$&$	0.83 	$&$	2.06 	$&$	0.64 	$&$	-0.16 	$&$	0.81 	$&$	2.05 	$\\
&&&&&&&&&&&&&& \\
	& &exp & CL& & & exp&\multicolumn{4}{c|}{CL} &\multicolumn{4}{l}{}  \\
$\CZ$ &  $I^\pi$ &$E_x$ &  $E_x$ &  & 
& $	\mu$&$\mu	$&$\langle S_z\rangle	$&$	\langle S^2\rangle	$&$		\langle L^2\rangle	$& & & & \\
$\CF$	&$	5^+	$&$	1.124 	$&$	0.94 	$&$		$&$		$&$	2.86(3)	$&$	2.88 	$&$	0.99 	$&$	2.00 	$&$	20.00 	$&$		$&$		$&$		$&$		$\\	
$\CF$	&$	3^+	$&$	0.937 	$&$	0.92 	$&$		$&$		$&$	1.77(12)	$&$	1.85 	$&$	0.93 	$&$	1.95 	$&$	6.54 	$&$		$&$		$&$		$&$		$\\	
$\CF$	&$	1^+_\textrm{gs}	$&$		$&$		$&$		$&$		$&$	-	$&$	0.81 	$&$	0.81 	$&$	1.94 	$&$	6.00 	$&$		$&$		$&$		$&$		$\\	
\hline
\end{tabular}
\end{center}
\end{table*}

\begin{table}[ht]
\caption{Proton radii in $\CZ$ nuclei obtained by the CL and AMD calculations together with 
the experimental data. 
The experimental values of proton radii  are reduced from charge radii \cite{Angeli13}.
\label{tab:radii}
}
\begin{center}
\begin{tabular}{cccccccccccccc}
\hline	
$\CZ$	& CL	&	AMD	&	exp	\\
 &  $R_p$ [fm]	&	$R_p$ [fm]	&	$R_p$ [fm]	\\
$^6\CLi$	&		2.56 	&	2.17 	&	2.452 	\\
$^7\CLi$	&	2.43 	&	2.16 	&	2.307 	\\
$^8\CBe$	&		3.37 	&	2.42 	&	$-$	\\
$^9\CBe$	&	2.60 	&	2.37 	&	2.384 	\\
$^9\CB$	&	2.86 	&	2.52 	&	$-$	\\
$^{10}\CB$	&		2.39 	&	2.26 	&	2.281 	\\
$^{11}\CB$	&	2.30 	&	2.18 	&	2.263 	\\
$^{11}\CC$	&	2.36 	&	2.24 	&	$-$	\\
$^{12}\CC$	&	2.35 	&	2.16 	&	2.326 	\\
$\CO$	&	2.58 	&	2.35 	&	$-$	\\
$\CF$	&	2.74 	&	$-$	&	$-$	\\
\hline			
\end{tabular}
\end{center}
\end{table}

The excitation energies calculated with the CL, AMD, AMD' and experimental values are summarized in Table \ref{tab:mu}. 
The calculated energy spectra depend on the
adopted $NN$ interactions as well as the structure models.
The CL result reasonably describes the low-lying energy spectra of $p$-shell nuclei.
The AMD generally gives similar results to the CL, but there are 
some exceptions. For example,  excitation energies of $^{10}\CB(1^+)$ and  $\CO(3/2^-)$ states 
are much overshot by the AMD calculation. The excitation energies of these states strongly 
depend on the strength of the $NN$ spin-orbit interactions. 
The overshooting is improved in the AMD' calculation because 
of  the weaker $NN$ spin-orbit interactions in the 
NN-a' and NN-c than that in the NN-a.

Table \ref{tab:radii} shows the comparison of 
root-mean-square (rms) radii ($R_p$) of proton distribution obtained by the 
CL and AMD calculations as well as the experimental values reduced from 
 rms charge radii. The AMD calculation generally gives smaller $R_p$ values 
than the CL calculation because the dynamical inter-cluster motion is not sufficiently 
taken into account in the present AMD. 
On the other hand, the CL model tends to give larger $R_p$ than the AMD because the cluster breaking is not taken into account in the model.
This is the major reason why we adopt the two typical nuclear models, CL and AMD, in this paper.
The experimental values are found to be in between theoretical values of two calculations.

\subsection{Spin configurations and magnetic moments of core nuclei  $\CZ$}
Spin configurations are not so sensitive to
choice of $NN$ interactions as the energy spectra
except for $p_{3/2}$-shell closed nuclei such as 
 $^{10}\CBe$, $^{11}$\CB, $^{11}\CC$, and $^{12}\CC$. 

In $\LZ$ with odd-odd and even-odd (odd-even) core nuclei, 
the spin-doublet splitting sensitively reflects the $z$ component 
$S_{z}$ of the total nuclear intrinsic-spin $\bvec{S}$ 
through the $\bvec{\sigma}_\La\cdot\bvec{\sigma}_N$ term
in the $\Delta_\sigma$ contribution.
It is able to check reliability of structure models for spin properties in comparison of 
$\mu$ moments in $\CZ$ between theory and experiment. 
The AMD and AMD' results for $\mu$ moments and 
nuclear intrinsic-spin and orbital angler momentum in 
nuclear states ($\CZ$)
are shown in Table \ref{tab:mu} together with the experimental $\mu$ moments.
The calculation reasonably reproduces the experimental $\mu$ moments
indicating that spin configurations of core nuclei are reasonably described with the AMD model.

In $Z=N$ odd-odd nuclei, $^6\CLi$, $^{10}\CB$, and $\CF$, 
$I^\pi=1^+$ and $I^\pi=3^+$ states are dominantly described by 
$pn$ pairs in $L=I-1$ states with the intrinsic spin 
$S=1$, which is approximately aligned to the total nuclear 
spin ($I$) as indicated by $\langle S_z\rangle \approx 1$. 
In particular, in $^6\CLi$ and $\CF$ states, 
$\alpha+pn$ and $^{16}\textrm{O}+pn$ structures are formed, respectively, and 
$\mu$ moments of $1^+$,  $3^+$, and $5^+$ states are close to the values $\mu=0.88$, 1.88 and 2.88 $\mu_N$ 
for the ideal $S=1$ $pn$ pairs in the $L=0$, 2, and 4 states, respectively. 

In even-odd and odd-even nuclei,  a valence nucleon spin $s=1/2$ tends to align to 
the total nuclear spin in $I^\pi=3/2^-$ states, but 
the alignment is not necessarily perfect because of significant configuration mixing as seen in deviation from $S_z=0.5$.
In $I^\pi=1/2^-$ states,
$S_z$ is roughly equal to $s_z=-1/6$ for the $p_{1/2}$ single-particle contribution, but 
non-negligible configuration mixing is contained as indicated by the $\mu$ moments 
slightly deviating from the Schmidt values ($\mu_\textrm{Shchmidt}=-0.79$ $\mu_N$ for a proton and 0.64 $\mu_N$ for 
a neutron)
except for $\CO(1/2^-)$. In $\CO(1/2^-)$, the spin configuration is understood by almost the pure $p_{1/2}$ hole configuration. 

In $Z=N=2n$ nuclei, $n\alpha$-cluster structures are favored. Particularly, 
the $^8\CBe(0^+,2^+)$ states have remarkable 2$\alpha$-cluster structures and 
almost pure $S=0$ components.
On the other hand, the $^{12}\CC(0^+)$ and  $^{12}\CC(2^+)$ states contain significant $S\ne 0$ components  
because of $3\alpha$-cluster breaking. The $S\ne 0$ mixing 
is sensitive to the $NN$ spin-orbit interactions especially in $^{12}\CC(0^+)$.  The mixing in $^{12}\CC(0^+)$ is less 
in the AMD' than the AMD because of the weaker $NN$ spin-orbit interaction. Compared with $^{12}\CC(0^+)$, 
the $S\ne 0$ component in $^{12}\CC(2^+)$ is 
relatively small and not so 
sensitive to the $NN$ interactions, but it still gives non-negligible $\Delta_\sigma$ 
and $T$ contributions to the spin-doublet splitting in $^{13}\LC$ spectra as discussed later. 

\section{Results of $\La$ hypernuclei: $\LZ$} \label{sec:results}

\subsection{Averaged structure properties of $\LZ$ states obtained with spin-independent $\La N$ interactions}

In this subsection, we show calculated results of averaged structure properties of
$\LZ$ states obtained with the leading part $V_0$  (spin-independent) of the $\La N$ interactions
without the perturbative part $V_1$. 


Table \ref{tab:BE-radii} shows the calculated results of 
the $\La$ binding energies $(\bar{B}_{\La,0})$,  rms radii of the $\La$ and nuclear distributions ($r_\La$ and $R_N$), and 
averaged Fermi momentum ($\langle k_F \rangle_\Lambda$) 
obtained with the CL and AMD. The experimental data of the 
the $\La$ binding energy $(B_\La)$ and spin-averaged one ($\bar{B}_\La$) are shown for comparison.
The systematics of the observed $\La$ binding energies is 
reasonably described by the leading part $V_0$ of
ECS08a(Hyb) interaction. 
The model dependence of $\bar{B}_{\La,0}$ between the CL and AMD  is not so strong. 
Quantitatively, 
the $\La$ binding is slightly deeper in the AMD than the CL except for $^{12}\LB(3/2^-_\textrm{gs})$, 
$^{12}\LC(3/2^-_\textrm{gs})$, and $^{13}\LC(0^+)$,
because the AMD tends to give smaller nuclear radii $R_N$, i.e., the higher nuclear density contributing the deeper 
$\La$-core potential.

\begin{figure}[!h]
\begin{center}
\includegraphics[width=8.6cm]{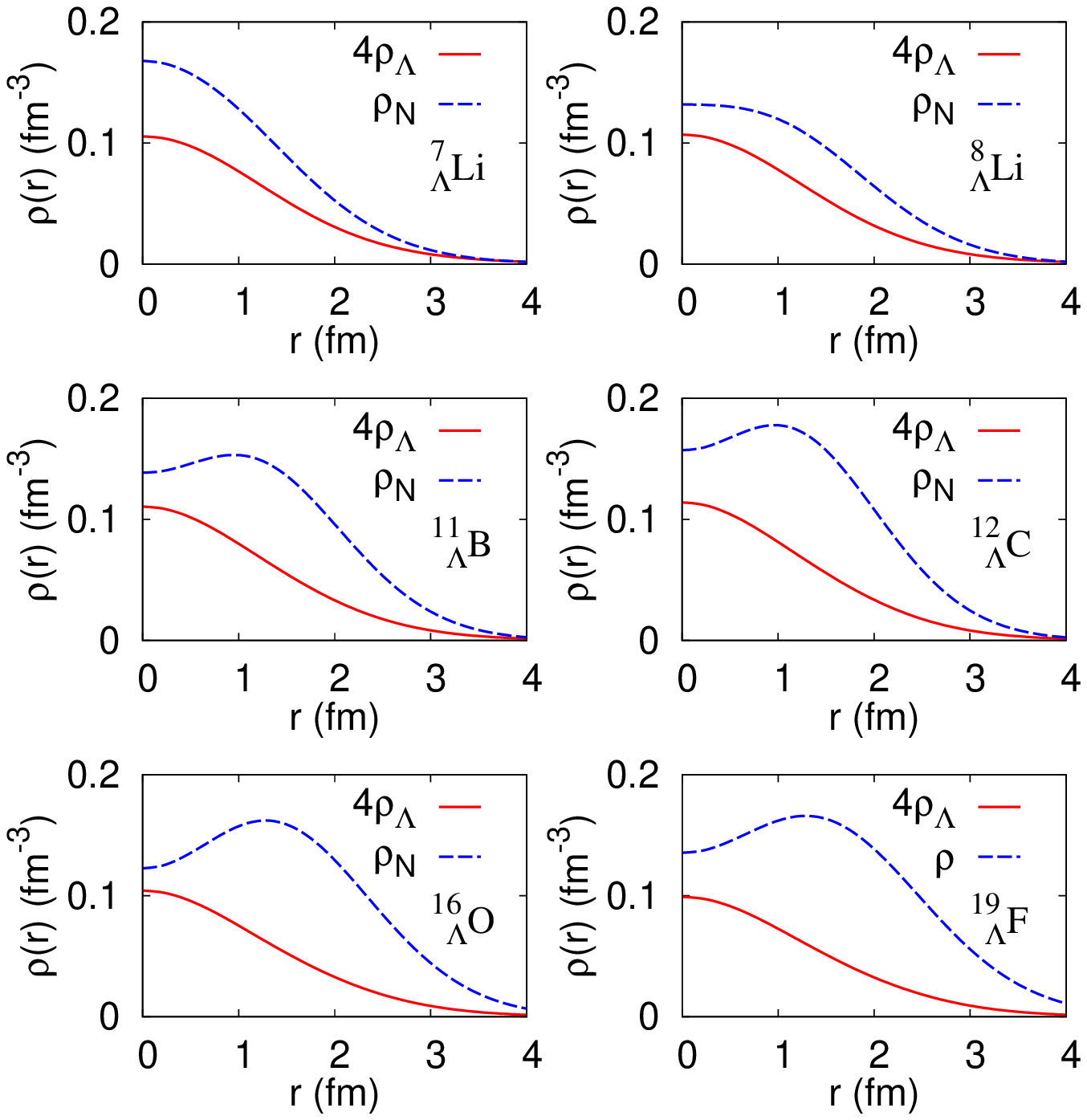} 	
\end{center}
  \caption{(color online) 
Distributions of the $\La$ ($\rho_\La(r)$) and nuclear matter $(\rho_N(r))$ densities 
 in $^{7}\LLi$, 
 $^{8}\LLi$,  $^{11}\LB$,  $^{12}\LC$,  $\LO$, and $\LF$ 
obtained with the CL calculation using the spin-independent $\La N$ interactions ($V_0$).
\label{fig:rho}}
\end{figure}

The $\La$ distribution size $(r_\La)$ is larger than the nuclear matter distribution size $(R_N)$
in light-mass $\LZ$  because of small $\La$ binding energies. 
With increase of the mass number $A$, $r_\La$ becomes gradually small as the $\La$ binding becomes
deep. 
The nuclear matter radii $R_N$  
increase with the increase of $A$, and in heavy-mass $p$-shell $\LZ$ it  is eventually as large as
$r_\La$.
Densities of $\La$ and nuclear distributions are shown in Fig.~\ref{fig:rho}. The mass number dependence of the 
$\La$ distribution is very mild compared with that of nuclear matter distributions.

\begin{table*}[ht]
\caption{The averaged $\La$ binging energies ($\bar{B}_{\La,0}$ [MeV]), rms radii  of the $\La$ distribution  ($r_\La$ [fm]), mean $k_F$ values 
($\langle k_F \rangle_\La$ [fm$^{-1}$]) , and rms radii of nuclear matter distribution ($R_N$ [fm]) in $\LZ$ for core $\CZ(I^\pi)$ states. 
The results obtained by the 
CL and AMD calculations with the spin-independent $\La N$ interactions ($V_0$)  are shown. The experimental 
$\La$ binding energies ($B_\La$ [MeV])  
\cite{Hashimoto:2006aw,Juric:1973zq,Davis:1992dt,Davis:2005mb,Ajimura:1998sy,Tang:2014atx}
 and the spin-averaged values ($\bar{B}_\La$ [MeV]) 
determined from spectroscopic studies \cite{Tamura:2010zz,Tang:2014atx} are also listed. 
\label{tab:BE-radii}
}
\begin{center}
\begin{tabular}{cc|cccc|cccc|cccc}
\hline			
$\LZ$ &($I^\pi$)& \multicolumn{4}{c|}{CL}& \multicolumn{4}{c|}{AMD}& \multicolumn{2}{c}{exp}\\
&&$	\bar{B}_{\La,0}	$&$	r_\La	$&$\langle k_F \rangle_\Lambda$&$	R_N	$&$	\bar{B}_{\La,0}	$&$	r_\La	$&$	\langle k_F \rangle_\Lambda$&$	R_N	$&$	B_\La$ &$\bar{B}_\La$ \\
$^7\LLi$	&$	(3^+)	$&$	5.93 	$&$	2.54 	$&$	0.98 	$&$	2.13 	$&$	5.96 	$&$	2.57 	$&$	0.98 	$&$	2.06 	$&$	-	$&$	-	$\\
$^7\LLi$	&$	(1^+_\textrm{gs})	$&$	5.35 	$&$	2.64 	$&$	0.93 	$&$	2.32 	$&$	5.66 	$&$	2.63 	$&$	0.94 	$&$	2.17 	$&$	5.58(3)	$&$	5.12(3)	$\\
&&&&&&&&&&&&&\\
$^8\LLi$	&$	(7/2^-)	$&$	7.02 	$&$	2.48 	$&$	0.99 	$&$	2.25 	$&$	6.88 	$&$	2.53 	$&$	0.99 	$&$	2.21 	$&$	-	$&$	-	$\\
$^8\LLi$	&$	(5/2^-)	$&$	6.15 	$&$	2.56 	$&$	0.95 	$&$	2.39 	$&$	6.74 	$&$	2.56 	$&$	0.97 	$&$	2.25 	$&$	-	$&$	-	$\\
$^8\LLi$	&$	(3/2^-_\textrm{gs})	$&$	6.68 	$&$	2.55 	$&$	0.96 	$&$	2.34 	$&$	6.73 	$&$	2.56 	$&$	0.97 	$&$	2.25 	$&$	-	$&$	-	$\\
$^8\LLi$	&$	(1/2^-)	$&$	6.48 	$&$	2.58 	$&$	0.94 	$&$	2.39 	$&$	6.68 	$&$	2.57 	$&$	0.97 	$&$	2.26 	$&$	-	$&$	-	$\\
&&&&&&&&&&&&&\\
$^9\LBe$	&$	(2^+)	$&$	5.15 	$&$	2.58 	$&$	0.94 	$&$	2.59 	$&$	7.41 	$&$	2.58 	$&$	0.95 	$&$	2.42 	$&$	-	$&$	-	$\\
$^9\LBe$	&$	(0^+_\textrm{gs})	$&$	6.53 	$&$	2.59 	$&$	0.93 	$&$	2.57 	$&$	7.42 	$&$	2.57 	$&$	0.96 	$&$	2.41 	$&$	6.71(4)	$&$	-	$\\
&&&&&&&&&&&&&\\
$^{10}\LBe$	&$	(5/2^-)	$&$	7.96 	$&$	2.52 	$&$	0.99 	$&$	2.55 	$&$	8.15 	$&$	2.55 	$&$	0.98 	$&$	2.52 	$&$-$&$	-	$\\
$^{10}\LBe$	&$	(3/2^-_\textrm{gs})	$&$	8.06 	$&$	2.51 	$&$	0.99 	$&$	2.54 	$&$	8.20 	$&$	2.54 	$&$	0.98 	$&$	2.50 	$&$	8.55(18)	$&$	-	$\\
$^{10}\LBe$	&$	(1/2^-)	$&$	7.33 	$&$	2.60 	$&$	0.95 	$&$	2.71 	$&$	7.80 	$&$	2.60 	$&$	0.95 	$&$	2.60 	$&$	-	$&$	-	$\\
&&&&&&&&&&&&&\\
$^{11}\LBe$	&$	(2^+)	$&$	9.10 	$&$	2.45 	$&$	1.07 	$&$	2.37 	$&$	9.02 	$&$	2.45 	$&$	1.10 	$&$	2.22 	$&$	-	$&$	-	$\\
$^{11}\LBe$	&$	(0^+_\textrm{gs})	$&$	9.01 	$&$	2.47 	$&$	1.06 	$&$	2.39 	$&$	9.04 	$&$	2.45 	$&$	1.10 	$&$	2.22 	$&$	-	$&$	-	$\\
&&&&&&&&&&&&&\\
$^{11}\LB$	&$	(3^+_\textrm{gs})	$&$	9.31 	$&$	2.44 	$&$	1.08 	$&$	2.34 	$&$	9.34 	$&$	2.43 	$&$	1.09 	$&$	2.26 	$&$	10.24(5)	$&$	10.09(5)	$\\
$^{11}\LB$	&$	(1^+)	$&$	8.54 	$&$	2.53 	$&$	1.01 	$&$	2.53 	$&$	8.71 	$&$	2.52 	$&$	1.03 	$&$	2.44 	$&$	-	$&$	-	$\\
&&&&&&&&&&&&&\\
$^{12}\LB$	&$	(5/2^-)	$&$	9.46 	$&$	2.47 	$&$	1.07 	$&$	2.45 	$&$	9.67 	$&$	2.43 	$&$	1.13 	$&$	2.25 	$&$	-	$&$	-	$\\
$^{12}\LB$	&$	(3/2^-_\textrm{gs})	$&$	10.06 	$&$	2.40 	$&$	1.13 	$&$	2.29 	$&$	9.76 	$&$	2.42 	$&$	1.14 	$&$	2.21 	$&$	11.38(2)	$&$	11.28(2)	$\\
$^{12}\LB$	&$	(1/2^-)	$&$	9.49 	$&$	2.47 	$&$	1.07 	$&$	2.45 	$&$	9.61 	$&$	2.44 	$&$	1.12 	$&$	2.28 	$&$	-	$&$	-	$\\
$^{12}\LB$	&$	(3/2^-_2)	$&$	9.21 	$&$	2.50 	$&$	1.05 	$&$	2.51 	$&$	9.55 	$&$	2.45 	$&$	1.11 	$&$	2.30 	$&$	-	$&$	-	$\\
&&&&&&&&&&&&&\\
$^{12}\LC$	&$	(5/2^-)	$&$	9.51 	$&$	2.47 	$&$	1.07 	$&$	2.46 	$&$	9.65 	$&$	2.44 	$&$	1.12 	$&$	2.26 	$&$	-	$&$	-	$\\
$^{12}\LC$	&$	(3/2^-_\textrm{gs})	$&$	10.14 	$&$	2.40 	$&$	1.13 	$&$	2.30 	$&$	9.73 	$&$	2.42 	$&$	1.14 	$&$	2.22 	$&$	10.76(19)	$&$	10.65(19)	$\\
$^{12}\LC$	&$	(1/2^-)	$&$	9.55 	$&$	2.47 	$&$	1.07 	$&$	2.46 	$&$	9.58 	$&$	2.45 	$&$	1.11 	$&$	2.29 	$&$	-	$&$	-	$\\
$^{12}\LC$	&$	(3/2^-_2)	$&$	9.25 	$&$	2.50 	$&$	1.05 	$&$	2.53 	$&$	9.53 	$&$	2.46 	$&$	1.11 	$&$	2.31 	$&$	-	$&$	-	$\\
&&&&&&&&&&&&&\\
$^{13}\LC$	&$	(2^+)	$&$	10.03 	$&$	2.46 	$&$	1.10 	$&$	2.44 	$&$	9.94 	$&$	2.43 	$&$	1.17 	$&$	2.21 	$&$	-	$&$	-	$\\
$^{13}\LC$	&$	(0^+_\textrm{gs})	$&$	10.44 	$&$	2.41 	$&$	1.15 	$&$	2.31 	$&$	9.99 	$&$	2.42 	$&$	1.19 	$&$	2.16 	$&$	11.69(12)	$&$-	$\\
&&&&&&&&&&&&&\\
$\LO$	&$	(3/2^-)	$&$	11.53 	$&$	2.47 	$&$	1.13 	$&$	2.61 	$&$	11.48 	$&$	2.43 	$&$	1.20 	$&$	2.38 	$&$	-	$&$	-	$\\
$\LO$	&$	(1/2^-_\textrm{gs})	$&$	11.71 	$&$	2.45 	$&$	1.15 	$&$	2.57 	$&$	11.57 	$&$	2.42 	$&$	1.22 	$&$	2.32 	$&$	12.42(5)	$&$	12.42(5)	$\\
&&&&&&&&&&&&&\\
$\LF$	&$	(5^+)	$&$	12.65 	$&$	2.48 	$&$	1.17 	$&$	2.70 	$&$	-	$&$	-	$&$	-	$&$	-	$&$	-	$&$	-	$\\
$\LF$	&$	(3^+)	$&$	12.70 	$&$	2.47 	$&$	1.18 	$&$	2.73 	$&$	-	$&$	-	$&$	-	$&$	-	$&$	-	$&$	-	$\\
$\LF$	&$	(1^+_\textrm{gs})	$&$	12.78 	$&$	2.47 	$&$	1.17 	$&$	2.74 	$&$	-	$&$	-	$&$	-	$&$	-	$&$	-	$&$	-	$\\
\hline			
\end{tabular}
\end{center}
\end{table*}


\subsection{Tuning of spin-dependence of $\La N$ interactions}


The $\Delta_\sigma$ term of the  ESC08(a,b) $\La NG$ interaction is known to be 
inappropriate to describe the observed spin-doublet splitting energies. 
For example, the ESC08a  gives 
the reverse ordering of the $J_>$ and $J_<$ states in $^{12}\LC$ and $^{11}\LB$ for 
the ground state core nuclei inconsistently with the experimental observation 
as pointed out in Ref.~\cite{Yamamoto:2010zzn}.
Other spin-dependent terms of the ESC08a $\La NG$ interaction have not been well tested yet.
In the present work, we phenomenologically tune the spin-dependent terms of the 
ESC08a $\La NG$ interaction by modifying the original 
strength parameters
to describe energy spectra in $p$-shell 
$\La$ hypernuclei as follows.

\begin{itemize}
\item
For the spin-independent term $V_0$, the original parameters are used.   
\item 
The $\Delta_\sigma$ term ($V_\sigma$) is adjusted so as to globally describe the 
$3/2^-$-$1/2^-$ and $7/2^-$-$5/2^-$ splittings 
in $^7\LLi$ and $^{11}\LB$, which are dominantly contributed by the $\Delta_\sigma$
term because of the total nuclear intrinsic-spin $S=1$ component. 
The even and odd parts of $V_\sigma$ are multiplied by factors of 2 and 0.3, respectively. 
\item
For the $T$ term ($V_T$), the observed value of the $1^-$-$0^-$ splitting
in $\LO$ is used as an input parameter for tuning.
In the $1^-$-$0^-$ splitting, the $T$ contribution is relatively large compared with 
other systems and cancels the $\Delta_\sigma$ contribution. 
The strength of $V_T$ is multiplied by a factor of 6 to fit the small $1^-$-$0^-$ splitting 
observed in $\LO$.
\item
The spin-orbit terms ($V_{S_\La}$ and $V_{S_N}$) are not modified. 
These contributions in splitting energies are generally small and it is difficult 
to definitely determine these terms without ambiguity from existing data,
and therefore, these terms are tentatively left as they are. 
However, it is likely that 
a larger $V_{S_\La}$ term with a factor of $\sim 2$ than the original one 
is favored to reproduce the $5/2^+$-$3/2^+$ splitting in $^9\LBe$. 
\end{itemize}

As a result, above modifications $v(r)\to \tilde v(r)$ are expressed as
\begin{eqnarray}
 \tilde v^{\textrm{e,o}}_0(r)&=&v^\textrm{e,o}_0(r), \\
 \tilde v^{\textrm{e,o}}_{\sigma}(r)&=&f^\textrm{e,o}_{\sigma} v^\textrm{e,o}_{\sigma}(r),\\
 \tilde v^{\textrm{o}}_{S_\La}(r)&=&f_{S_\La} v^\textrm{o}_{S_\La}(r),\\
 \tilde v^{\textrm{o}}_{S_N}(r)&=&f_{S_N} v^\textrm{o}_{S_N}(r),\\
 \tilde v^{\textrm{o}}_{T}(r)&=&f_{T} v^\textrm{o}_{T}(r),
\end{eqnarray}
with $f^\textrm{e}_\sigma=2$, $f^\textrm{o}_\sigma=0.3$,  $f_{S_\La}=f_{S_N}=1$, and $f_T=6$. 
We call the ESC08a $\La NG$ interaction with thus modified spin dependence (msd) ``ESC08a-msd''.
In the following sections, we first show the crucial problem of the original spin-dependent interactions of 
ESC08a in reproducing the experimental spin-doublet splitting energies and 
how the results are improved by the modified interactions, ESC08a-msd. Then, we discuss 
the details of energy spectra in $\LZ$ based on the calculations with the ESC08a-msd, which 
we use as the default $\La N$ interactions unless otherwise specified.

\subsection{Spin-doublet splitting energies: general features}

The spin-doublet splitting energies and respective contributions calculated with the original ESC08a and ESC08a-msd 
interactions 
are shown in Table \ref{tab:splitting-orig} together with the experimental data. 
Significant splittings have been experimentally observed for such core nuclear states as $\CZ(1^+_1)$ and $\CZ(3^+_1)$ in
$Z=N$ odd-odd nuclei and  $\CZ(3/2^-_\textrm{gs})$ and $\CZ(5/2^-_1)$ in even-odd and odd-even core nuclei. 
However, the original ESC08a $\La NG$ interaction gives opposite-sign splitting energies, namely, 
the reverse ordering of the spin-doublet partners 
because opposite-sign contributions from the odd term of $V_\sigma$ dominate the $\Delta_\sigma$ contributions.
It is a crucial problem of the spin dependence of the original ESC08a 
as pointed out in Ref.~\cite{Yamamoto:2010zzn}. 
In contrast, the spin-doublet splitting energies are reasonably reproduced by the ESC08a-msd with the 
modified spin-dependence of the $\La N$ interactions. 
The significant splittings observed for $\CZ(1^+_1)$ and $\CZ(3^+_1)$ of $Z=N$ odd-odd nuclei, 
and  $\CZ(3/2^-_\textrm{gs})$ and $\CZ(5/2^-_1)$ of even-odd and odd-even core nuclei 
are described by the dominant $\Delta_\sigma$ contributions in the present AMD calculation with the ESC08a-msd.
$T$ contributions are usually small compared with dominant $\Delta_\sigma$ contributions except for core
$\CZ(1/2^-)$ states in odd-even and even-odd nuclei, in which the $T$ and $\Delta_\sigma$ contributions are comparable order and 
almost cancel with each other. 
The $S_\La$ term gives minor contributions in general. 

\begin{table*}[ht]
\caption{Spin-doublet splittings in $\LZ$ for core $\CZ(I^\pi)$ states
obtained by the AMD calculation with the original (ESC08a) and 
modified (ESC08a-msd) spin dependences of the $\La N$ interactions.
The $\Delta_\sigma$, $S_\La$, $T$ contributions as well as the total splitting energies are listed. 
The experimental values of the splittings from Refs.\cite{Tamura:2010zz,Chrien:1990ag,Tanida:2000zs,Ukai:2006zp,Akikawa:2002tm,Ajimura:2001na,Miura:2005mh,Ma:2010zzb,Kohri:2001nc,Tang:2014atx,Hosomi:2015fma,Ukai:2008aa,Yang:2017lay}
 are also shown.
$^{a}$For the experimental value in $^8\LLi$,  the 442 keV $\gamma$-ray was tentatively attributed to the transitions in $^8\LLi$ and $^8\LBe$ \cite{Chrien:1990ag}.
\label{tab:splitting-orig}
}
\begin{center}
\begin{tabular}{cccc|cccc|cc|cccc}
\hline			
	$\LZ$ &($I^\pi$) &	$J^\pi_>$ 	&$J^\pi_<$ 		&	\multicolumn{4}{|c|}{Modified (ESC08a-msd)}	&	\multicolumn{2}{c|}{original (ESC08a)}
&	exp  \\
	&	&	&		&	$\Delta_\sigma$	&	$S_\La$	&	$T $	&	Total	& $\Delta_\sigma$		&	Total	
&	 \\
$^7\LLi$ & $(3^+)$	&$	7/2^+	$&$	5/2^+	$&$	0.474 	$&$	-0.018 	$&$	-0.061 	$&$	0.396 	$&$	-0.399 	$&$	-0.427 	$&	0.471	\\
$^7\LLi$	 & $(1^+_\textrm{gs})$ &$	3/2^+	$&$	1/2^+	$&$	0.522 	$&$	0.000 	$&$	0.001 	$&$	0.523 	$&$	-0.423 	$&$	-0.422 	$&	0.692	\\
&&&&&&&&&&&&&\\
$^8\LLi$ & $(7/2^-)$	&$	4^-	$&$	3^-	$&$	0.291 	$&$	-0.026 	$&$	-0.045 	$&$	0.219 	$&$	-0.195 	$&$	-0.229 	$&	-	\\
$^8\LLi$ & $(5/2^-)$	&$	3^-	$&$	2^-	$&$	-0.126 	$&$	-0.026 	$&$	0.041 	$&$	-0.111 	$&$	0.136 	$&$	0.117 	$&	-	\\
$^8\LLi$ & $(3/2^-_\textrm{gs})$	&$	2^-	$&$	1^-	$&$	0.286 	$&$	-0.010 	$&$	-0.024 	$&$	0.252 	$&$	-0.204 	$&$	-0.218 	$&	$0.442^{a}$	\\
$^8\LLi$ & $(1/2^-)$	&$	1^-	$&$	0^-	$&$	-0.117 	$&$	-0.010 	$&$	0.095 	$&$	-0.032 	$&$	0.089 	$&$	0.095 	$&	-	\\
&&&&&&&&&&&&&\\
$^9\LBe$ & $(2^+)$	&$	5/2^+	$&$	3/2^+	$&$	-0.002 	$&$	-0.018 	$&$	-0.003 	$&$	-0.023 	$&$	-0.007 	$&$	-0.025 	$&	-0.043	\\
&&&&&&&&&&&&&\\
$^{10}\LBe$ & $(5/2^-)$	&$	3^-	$&$	2^-	$&$	0.206 	$&$	-0.020 	$&$	0.010 	$&$	0.196 	$&$	-0.106 	$&$	-0.124 	$&	-	\\
$^{10}\LBe$ & $(3/2^-_\textrm{gs})$	&$	2^-	$&$	1^-	$&$	0.192 	$&$	-0.013 	$&$	-0.021 	$&$	0.159 	$&$	-0.106 	$&$	-0.122 	$&	-	\\
$^{10}\LBe$ & $(1/2^-)$	&$	1^-	$&$	0^-	$&$	-0.095 	$&$	-0.011 	$&$	0.099 	$&$	-0.007 	$&$	0.045 	$&$	0.051 	$&	-	\\
&&&&&&&&&&&&&\\
$^{11}\LBe$ & $(2^+)$	&$	5/2^+	$&$	3/2^+	$&$	0.098 	$&$	-0.018 	$&$	0.004 	$&$	0.084 	$&$	-0.068 	$&$	-0.086 	$&	-	\\
&&&&&&&&&&&&&\\
$^{11}\LB$	 & $(3^+_\textrm{gs})$ &$	7/2^+	$&$	5/2^+	$&$	0.487 	$&$	-0.020 	$&$	-0.060 	$&$	0.407 	$&$	-0.374 	$&$	-0.404 	$&	0.264	\\
$^{11}\LB$ & $(1^+)$	&$	3/2^+	$&$	1/2^+	$&$	0.408 	$&$	-0.004 	$&$	-0.044 	$&$	0.360 	$&$	-0.253 	$&$	-0.264 	$&	0.505	\\
&&&&&&&&&&&&&\\
$^{12}\LB$ & $(5/2^-)$	&$	3^-	$&$	2^-	$&$	0.313 	$&$	-0.020 	$&$	0.059 	$&$	0.352 	$&$	-0.225 	$&$	-0.235 	$&	-	\\
$^{12}\LB$ & $(3/2^-_\textrm{gs})$	&$	2^-	$&$	1^-	$&$	0.172 	$&$	-0.014 	$&$	-0.055 	$&$	0.103 	$&$	-0.129 	$&$	-0.152 	$&	0.179	\\
$^{12}\LB$ & $(1/2^-)$	&$	1^-	$&$	0^-	$&$	-0.130 	$&$	-0.011 	$&$	0.131 	$&$	-0.010 	$&$	0.096 	$&$	0.106 	$&	-	\\
$^{12}\LB$ & $(3/2^-_2)$	&$	2^-	$&$	1^-	$&$	0.021 	$&$	-0.017 	$&$	-0.049 	$&$	-0.045 	$&$	-0.012 	$&$	-0.037 	$&	-	\\
&&&&&&&&&&&&&\\
$^{12}\LC$ & $(5/2^-)$	&$	3^-	$&$	2^-	$&$	0.313 	$&$	-0.020 	$&$	0.059 	$&$	0.352 	$&$	-0.225 	$&$	-0.235 	$&	-	\\
$^{12}\LC$ & $(3/2^-_\textrm{gs})$	&$	2^-	$&$	1^-	$&$	0.175 	$&$	-0.014 	$&$	-0.055 	$&$	0.106 	$&$	-0.132 	$&$	-0.155 	$&	0.161	\\
$^{12}\LC$ & $(1/2^-)$	&$	1^-	$&$	0^-	$&$	-0.131 	$&$	-0.011 	$&$	0.130 	$&$	-0.012 	$&$	0.096 	$&$	0.107 	$&	-	\\
$^{12}\LC$ & $(3/2^-_2)$	&$	2^-	$&$	1^-	$&$	0.018 	$&$	-0.017 	$&$	-0.048 	$&$	-0.047 	$&$	-0.010 	$&$	-0.035 	$&	-	\\
&&&&&&&&&&&&&\\
$^{13}\LC$ & $(2^+)$	&$	5/2^+	$&$	3/2^+	$&$	0.049 	$&$	-0.019 	$&$	0.034 	$&$	0.064 	$&$	-0.070 	$&$	-0.083 	$&	-	\\
&&&&&&&&&&&&&\\
$\LO$ & $(3/2^-)$	&$	2^-	$&$	1^-	$&$	0.328 	$&$	-0.012 	$&$	-0.034 	$&$	0.282 	$&$	-0.147 	$&$	-0.164 	$&	0.224	\\
$\LO$ & $(1/2^-_\textrm{gs})$	&$	1^-	$&$	0^-	$&$	-0.167 	$&$	-0.011 	$&$	0.184 	$&$	0.006 	$&$	0.099 	$&$	0.119 	$&	0.026	\\
&&&&&&&&&&&&&\\
$\LF$ & $(5^+)$	&$	11/2^+	$&$	9/2^+	$&$	0.199 	$&$	-0.019 	$&$	-0.052 	$&$	0.128 	$&$	-0.372 	$&$	-0.399 	$&	-	\\
$\LF$ & $(3^+)$	&$	7/2^+	$&$	5/2^+	$&$	0.219 	$&$	-0.011 	$&$	-0.031 	$&$	0.177 	$&$	-0.328 	$&$	-0.344 	$&	-	\\
$\LF$ & $(1^+_\textrm{gs})$	&$	3/2^+	$&$	1/2^+	$&$	0.219 	$&$	0.000 	$&$	0.031 	$&$	0.249 	$&$	-0.356 	$&$	-0.351 	$&	0.315	\\
\hline
\end{tabular}
\end{center}
\end{table*}

The present result of splitting energies is compared with SM calculations in  Table \ref{tab:splitting-SM}.
Each contribution is also compared with that of the 
SM calculation by Millener {\it et al.} \cite{Millener:2012zz}.  
In general, the $\Delta_\sigma$, $S_\La$, and $T$ contributions obtained in the present calculation are 
similar to those of the Millener's SM calculation. 
This is a natural consequence because, in both calculations, spin-dependent contributions are 
phenomenologically adjusted to fit the observed splitting energies in $p$-shell $\LZ$. 

The present AMD calculation with the ESC08a-msd interaction, reasonably describes 
the global feature of  observed data of spin-doublet splittings in $p$-shell $\LZ$ though agreements with the experimental value are 
not so precise as the Millener's SM calculation \cite{Millener:2012zz}. 
It should be commented that,  in the Millener's SM calculation, 
the spin-dependent $\La N$ interaction parameters are independently adjusted to 
the light- and heavy-mass $p$-shell regions, whereas in the present calculation
the mass-number independent $\La N$ interactions are used.

\begin{table*}[ht]
\caption{Spin-doublet splittings in $\LZ$ for core $\CZ(I^\pi)$ states obtained by the AMD calculation with the 
ESC08a-msd compared with the experimental and SM calculations. 
The $\Delta_\sigma$, $S_\La$, $T$ contributions, and total splittings are listed together with the Millener's
SM calculation \cite{Millener:2012zz}. 
$\La \Sigma$ and $S_N$ contributions in the Millener's  calculation are also shown. 
Other SM calculations are taken from Refs.~\cite{Umeya-10Be,Motoba:2017lkk,Umeya:2016bbt}.
Information of the experimental values is explained in the caption of Table \ref{tab:splitting-orig}.
\label{tab:splitting-SM}
}
\begin{center}
\begin{tabular}{cccc|cccc|cc|cccccc}
\hline	
$\LZ$ & ($I^\pi$) &	$J^\pi_>$ 	&$J^\pi_<$ 		&		\multicolumn{4}{c|}{Present}	&
exp	& SM & \multicolumn{6}{c}{SM\cite{Millener:2012zz}}\\
& &	&	&
$\Delta_\sigma$	&	$S_\La$	&	$T $	&	Total	& 	&Total&	Total & $\Delta_\sigma$	&	$S_\La$	& $T$ &   $\La\Sigma$ &$S_N$ \\
$^7\LLi$ & $(3^+)$	&$	7/2^+	$&$	5/2^+	$&$	0.474 	$&$	-0.018 	$&$	-0.061 	$&$	0.396 	$&	0.471	&$		$&$	0.494	$&$	0.557	$&$	-0.032	$&$	-0.071	$&$	0.074	$&$	-0.008	$\\
$^7\LLi$ & $(1^+_\textrm{gs})$	&$	3/2^+	$&$	1/2^+	$&$	0.522 	$&$	0.000 	$&$	0.001 	$&$	0.523 	$&	0.692	&$		$&$	0.693	$&$	0.628	$&$	-0.001	$&$	-0.009	$&$	0.072	$&$	-0.004	$\\
&&&&&&&&&&&&&&&\\
$^8\LLi$ & $(7/2^-)$	&$	4^-	$&$	3^-	$&$	0.291 	$&$	-0.026 	$&$	-0.045 	$&$	0.219 	$&	-	&$		$&$	0.307	$&$	-	$&$	-	$&$	-	$&$	-	$&$	-	$\\
$^8\LLi$ & $(3/2^-_\textrm{gs})$	&$	2^-	$&$	1^-	$&$	0.286 	$&$	-0.010 	$&$	-0.024 	$&$	0.252 	$&	$0.442^{a}$	&$		$&$	0.445	$&$	0.393 	$&$	-0.014 	$&$	-0.023 	$&$	0.149 	$&$	-0.015 	$\\
$^8\LLi$ & $(1/2^-)$	&$	1^-	$&$	0^-	$&$	-0.117 	$&$	-0.010 	$&$	0.095 	$&$	-0.032 	$&	-	&$		$&$	0.006	$&$	-	$&$	-	$&$	-	$&$	-	$&$	-	$\\
&&&&&&&&&&&&&&&\\
$^9\LBe$ & $(2^+)$	&$	5/2^+	$&$	3/2^+	$&$	-0.002 	$&$	-0.018 	$&$	-0.003 	$&$	-0.023 	$&	-0.043	&$		$&$	-0.044	$&$	0.014 	$&$	-0.037 	$&$	-0.028 	$&$	0.008 	$&$	0.000 	$\\
&&&&&&&&&&&&&&&\\
$^{10}\LBe$	 & $(5/2^-)$&$	3^-	$&$	2^-	$&$	0.206 	$&$	-0.020 	$&$	0.010 	$&$	0.196 	$&	-	&$	0.148	$\cite{Umeya-10Be}&$	0.103	$&$	0.172 	$&$	-0.037 	$&$	-0.010 	$&$	-0.019 	$&$	-0.005 	$\\
$^{10}\LBe$ & $(3/2^-_\textrm{gs})$	&$	2^-	$&$	1^-	$&$	0.192 	$&$	-0.013 	$&$	-0.021 	$&$	0.159 	$&	-	&$	0.165	$\cite{Umeya-10Be}&$	0.110	$&$	0.180 	$&$	-0.022 	$&$	-0.033 	$&$	-0.010 	$&$	-0.004 	$\\
$^{10}\LBe$	 & $(1/2^-)$&$	1^-	$&$	0^-	$&$	-0.095 	$&$	-0.011 	$&$	0.099 	$&$	-0.007 	$&	-	&$	-0.162	$\cite{Umeya-10Be}&$	0.026	$&$	-	$&$	-	$&$	-	$&$	-	$&$	-	$\\
&&&&&&&&&&&&&&&\\
$^{11}\LB$  & $(3^+_\textrm{gs})$	&$	7/2^+	$&$	5/2^+	$&$	0.487 	$&$	-0.020 	$&$	-0.060 	$&$	0.407 	$&	0.264	&$		$&$	0.267	$&$	0.339 	$&$	-0.037 	$&$	-0.080 	$&$	0.056 	$&$	-0.010 	$\\
$^{11}\LB$  & $(1^+)$	&$	3/2^+	$&$	1/2^+	$&$	0.408 	$&$	-0.004 	$&$	-0.044 	$&$	0.360 	$&	0.505	&$		$&$	0.475	$&$	0.424 	$&$	-0.003 	$&$	-0.010 	$&$	0.061 	$&$	-0.044 	$\\
&&&&&&&&&&&&&&&\\
$^{12}\LB$ & $(5/2^-)$	&$	3^-	$&$	2^-	$&$	0.313 	$&$	-0.020 	$&$	0.059 	$&$	0.352 	$&	-	&$	0.389	$\cite{Motoba:2017lkk}&$	-	$&$	-	$&$	-	$&$	-	$&$	-	$&$	-	$\\
$^{12}\LB$ & $(3/2^-_\textrm{gs})$	&$	2^-	$&$	1^-	$&$	0.172 	$&$	-0.014 	$&$	-0.055 	$&$	0.103 	$&	0.179 &$	0.186$\cite{Motoba:2017lkk}$	$&$	-	$&$	-	$&$	-	$&$	-	$&$	-	$&$	-	$\\
$^{12}\LB$ & $(1/2^-)$	&$	1^-	$&$	0^-	$&$	-0.130 	$&$	-0.011 	$&$	0.131 	$&$	-0.010 	$&	-	&$	-0.664$\cite{Motoba:2017lkk}&$	-	$&$	-	$&$	-	$&$	-	$&$	-	$&$	-	$\\
$^{12}\LB$ & $(3/2^-_2)$	&$	2^-	$&$	1^-	$&$	0.021 	$&$	-0.017 	$&$	-0.049 	$&$	-0.045 	$&	-	&$	-0.122 $\cite{Motoba:2017lkk}$	$&$	-	$&$	-	$&$	-	$&$	-	$&$	-	$&$	-	$\\
&&&&&&&&&&&&&&&\\
$^{12}\LC$ & $(3/2^-_\textrm{gs})$	&$	2^-	$&$	1^-	$&$	0.175 	$&$	-0.014 	$&$	-0.055 	$&$	0.106 	$&	0.161	&$		$&$	0.153	$&$	0.175 	$&$	-0.012 	$&$	-0.042 	$&$	0.061 	$&$	-0.013 	$\\
&&&&&&&&&&&&&&&\\
$\LO$ & $(3/2^-)$	&$	2^-	$&$	1^-	$&$	0.328 	$&$	-0.012 	$&$	-0.034 	$&$	0.282 	$&	0.224	&$		$&$	0.248	$&$	0.207 	$&$	-0.021 	$&$	-0.041 	$&$	0.092 	$&$	0.001 	$\\
$\LO$ & $(1/2^-_\textrm{gs})$	&$	1^-	$&$	0^-	$&$	-0.167 	$&$	-0.011 	$&$	0.184 	$&$	0.006 	$&	0.026	&$		$&$	0.023	$&$	-0.123 	$&$	-0.020 	$&$	0.188 	$&$	-0.033 	$&$	0.001 	$\\		
&&&&&&&&&&&&&&&\\
$\LF$ & $(5^+)$	&$	11/2^+	$&$	9/2^+	$&$	0.199 	$&$	-0.019 	$&$	-0.052 	$&$	0.128 	$&	-	&$	0.408$\cite{Umeya:2016bbt}$	$&$	-	$&$	-	$&$	-	$&$	-	$&$	-	$&$	-	$\\
$\LF$ & $(3^+)$	&$	7/2^+	$&$	5/2^+	$&$	0.219 	$&$	-0.011 	$&$	-0.031 	$&$	0.177 	$&	-	&$	0.562$\cite{Umeya:2016bbt}$	$&$	0.196	$&$	-	$&$	-	$&$	-	$&$	-	$&$	-	$\\
$\LF$ & $(1^+_\textrm{gs})$	&$	3/2^+	$&$	1/2^+	$&$	0.219 	$&$	0.000 	$&$	0.031 	$&$	0.249 	$&	0.315	&$	0.419$\cite{Umeya:2016bbt}$	$&$	0.305	$&$	-	$&$	-	$&$	-	$&$	-	$&$	-	$\\

\hline
\end{tabular}
\end{center}
\end{table*}	

Let us discuss the $NN$ interaction dependence of splitting energies.
In Table \ref{tab:nuclear-int-dep}, 
we compare the AMD result (the stronger $NN$ spin-orbit interaction)  
and the AMD' result (the weaker $NN$ spin-orbit interaction).
The difference of the $NN$ spin-orbit interaction causes slight difference in the nuclear intrinsic-spin configurations. 
Generally, weaker $NN$ spin-orbit interaction enhances the $LS$-coupling component and reduces the $jj$-coupling component (the cluster breaking).  
However, in most cases, the $NN$ interaction difference between the AMD' and AMD gives only minor difference
in the splittings because nuclear intrinsic-spin configurations are not so sensitive to the $NN$ interactions
as shown in Table \ref{tab:mu}.

\begin{table*}[ht]
\caption{$NN$ interaction dependence of the spin-doublet splittings in $\LZ$ for core $\CZ(I^\pi)$ states. The results of the AMD and AMD' 
calculations with the ESC8a-msd interactions are shown.
Information of the experimental values is explained in the caption of Table \ref{tab:splitting-orig}.
\label{tab:nuclear-int-dep}
 }
\begin{center}
\begin{tabular}{cccc|cccc|c|ccccc}
\hline	
$\LZ$ &	($I^\pi$) & $J^\pi_>$ 	&$J^\pi_<$ 		&	
\multicolumn{4}{c|}{AMD}	& exp& \multicolumn{4}{c}{AMD'}	\\
	&	&	&		&	$\Delta_\sigma$	&	$S_\La$	&	$T $	&	Total	& 	& $\Delta_\sigma$	&	$S_\La$	& $T$& Total \\
$^9\LBe$ & $(2^+)$	&$	5/2^+	$&$	3/2^+	$&$	-0.002 	$&$	-0.018 	$&$	-0.003 	$&$	-0.023 	$&	$-0.043$ 	&$	-0.002 	$&$	-0.018 	$&$	-0.003 	$&$	-0.023 	$\\
&&&&&&&&&&&&&\\
$^{10}\LBe$	 & $(5/2^-)$&$	3^-	$&$	2^-	$&$	0.206 	$&$	-0.020 	$&$	0.010 	$&$	0.196 	$&	-	&$	0.193 	$&$	-0.020 	$&$	0.015 	$&$	0.188 	$\\
$^{10}\LBe$	 & $(3/2^-_\textrm{gs})$&$	2^-	$&$	1^-	$&$	0.192 	$&$	-0.013 	$&$	-0.021 	$&$	0.159 	$&	-	&$	0.189 	$&$	-0.012 	$&$	-0.018 	$&$	0.159 	$\\
$^{10}\LBe$ & $(1/2^-)$	&$	1^-	$&$	0^-	$&$	-0.095 	$&$	-0.011 	$&$	0.099 	$&$	-0.007 	$&	-	&$	-0.095 	$&$	-0.011 	$&$	0.097 	$&$	-0.009 	$\\
&&&&&&&&&&&&&\\
$^{11}\LBe$	 & $(2^+)$ &$	5/2^+	$&$	3/2^+	$&$	0.098 	$&$	-0.018 	$&$	0.004 	$&$	0.084 	$&	-	&$	0.027 	$&$	-0.020 	$&$	0.013 	$&$	0.019 	$\\
&&&&&&&&&&&&&\\
$^{11}\LB$ & $(3^+_\textrm{gs})$	&$	7/2^+	$&$	5/2^+	$&$	0.487 	$&$	-0.020 	$&$	-0.060 	$&$	0.407 	$&	0.264 	&$	0.464 	$&$	-0.021 	$&$	-0.052 	$&$	0.391 	$\\
$^{11}\LB$ & $(1^+)$	&$	3/2^+	$&$	1/2^+	$&$	0.408 	$&$	-0.004 	$&$	-0.044 	$&$	0.360 	$&	0.505 	&$	0.433 	$&$	-0.004 	$&$	-0.043 	$&$	0.387 	$\\
&&&&&&&&&&&&&\\
$^{12}\LC$ & $(5/2^-)$	&$	3^-	$&$	2^-	$&$	0.313 	$&$	-0.020 	$&$	0.059 	$&$	0.352 	$&	-	&$	0.282 	$&$	-0.020 	$&$	0.058 	$&$	0.319 	$\\
$^{12}\LC$ & $(3/2^-_\textrm{gs})$	&$	2^-	$&$	1^-	$&$	0.175 	$&$	-0.014 	$&$	-0.055 	$&$	0.106 	$&	0.161 	&$	0.166 	$&$	-0.014 	$&$	-0.054 	$&$	0.098 	$\\
$^{12}\LC$ & $(1/2^-)$	&$	1^-	$&$	0^-	$&$	-0.131 	$&$	-0.011 	$&$	0.130 	$&$	-0.012 	$&	-	&$	-0.135 	$&$	-0.011 	$&$	0.128 	$&$	-0.018 	$\\
$^{12}\LC$ & $(3/2^-_2)$	&$	2^-	$&$	1^-	$&$	0.018 	$&$	-0.017 	$&$	-0.048 	$&$	-0.047 	$&	-	&$	-0.009 	$&$	-0.017 	$&$	-0.047 	$&$	-0.073 	$\\
&&&&&&&&&&&&&\\
$^{13}\LC$ & $(2^+)$	&$	5/2^+	$&$	3/2^+	$&$	0.049 	$&$	-0.019 	$&$	0.034 	$&$	0.064 	$&	-	&$	0.025 	$&$	-0.019 	$&$	0.020 	$&$	0.026 	$\\
\hline
\end{tabular}
\end{center}
\end{table*}

\subsection{Spin-doublet splitting energies: characteristics in cases of odd-odd, even-odd(odd-even), and even-even core nuclei}
We here discuss 
characteristics of splitting energies in cases of odd-odd, even-odd(odd-even), and even-even core nuclei
based on the AMD results with the ESC08a-msd in Table \ref{tab:splitting-orig} and 
the comparison with the SM calculations 
in Table \ref{tab:splitting-SM}. 
\subsubsection{Case of $Z=N$ odd-odd core nuclei: 
$3/2^+$-$1/2^+$  and $7/2^+$-$5/2^+$ splittings in $^7\LLi$, $^{11}\LB$, and $\LF$}

In the case of $\LZ$ with $Z=N$ odd-odd  core nuclei, one of the characteristics of $T=0$ states is remarkably 
large splitting energies
because of the dominant nuclear intrinsic-spin $S=1$ component contributed by 
two valence nucleons, a proton and a neutron.
As shown in Table \ref{tab:mu}, the core nuclear states, $I^\pi=1^+$, $3^+$, and $5^+$ 
have $S=1$ and $T=0$ $pn$ pairs in the $L=0$, 2, and 4 waves as dominant components, respectively.
The aligned nuclear intrinsic-spin $S_z\approx 1$ provides the significant $\Delta_\sigma$ contribution in the splitting.

The mass number dependence of $\Delta_\sigma$ comes from the difference in spatial overlap 
between valence nucleon and $\La$ orbits. In $\LF$,  the $0s$-orbit $\La$ has a 
smaller overlap with the $sd$-orbit proton and neutron than in $^7\LLi$ and $^{11}\LB$ having
the valence proton and neutron in the $p$-shell. 
In each $\LZ$ system, higher $J$ states tend to have smaller splittings because
of negative contributions from the $T$ and $S_\La$ terms. 

The splittings in $^7\LLi$ have been investigated in details by 
the OCM (semi-microscopic) cluster model calculation with the NSC97f 
$\La NG$ interactions \cite{Hiyama:2006xv}.
In the OCM calculations with the NSC97f, the $\Delta_\sigma$ contribution in  the 
$3/2^+$-$1/2^+$($7/2^+$-$5/2^+$) splitting is 
$\Delta_{\sigma}=0.71 (0.54)$ MeV with  
even and odd contributions, $\Delta^\textrm{e}_{\sigma}=1.10(1.04)$ MeV and 
$\Delta^\textrm{o}_{\sigma}=-0.39(-0.50)$ MeV, respectively. 
In the present AMD with the ESC08a-msd, it is $\Delta_{\sigma}=0.52 (0.47)$ MeV
 with $\Delta^\textrm{e}_{\sigma}=0.75 (0.72)$ MeV and 
$\Delta^\textrm{o}_{\sigma}=-0.22 (-0.25)$ MeV.
The present result is qualitatively consistent with  but quantitatively slightly smaller than the OCM calculation
and also slightly underestimates the experimental data.
For more precise reproduction of the splittings in $^7\LLi$, higher order effects
beyond the $s$-wave $\La$ approximation should be taken into account.

For $\LF$, the AMD calculation gives  the $3/2^+$-$1/2^+$ splitting of 0.249 MeV, which reasonably agrees with the 
experimental value 0.315 MeV recently observed by the $\gamma$-ray measurement \cite{Yang:2017lay}. 
For the $7/2^+$-$5/2+$ and $11/2^+$-$9/2+$ splittings in $\LF$, the present calculation predicts 
smaller splittings than the $3/2^+$-$1/2^+$ splitting 
because of the cancellation of the dominant $\Delta_\sigma$ contribution 
by opposite-sign $T$ and $S_\La$ contributions. Our result is consistent with Millener's SM prediction but 
seems appreciably different from the prediction estimated  
with NSC97f $G$-matrix interaction by Umeya et al. \cite{Umeya:2016bbt}.

\subsubsection{Case of even-odd and odd-even core nuclei: splittings in $^{8}\LLi$, $^{10}\LBe$, $^{12}\LB$, $^{12}\LC$, $\LO$}

In $\LZ$ with even-odd and odd-even core nuclei, the $2^-$-$1^-$ and $3^-$-$2^-$ splittings 
for core $I^\pi=3/2^-$ and $5/2^-$ states are moderate. They are 
mainly contributed by the valence nucleon spin $S=1/2$ in the $p$-shell through the $\Delta_\sigma$ term.
The exception is the splitting for the core $^7\CLi(5/2^-)$ state, which is described by the $t$ cluster orbiting 
in the $L=3$ wave around the $\alpha$ cluster. In the $I^\pi=5/2^-$ state, the $t$-cluster intrinsic-spin $S=1/2$
is coupling with the $L=3$ with the anti-parallel orientation, and 
gives a negative contribution to the splitting. As a result, the ordering of the $3^-$ and $2^-$ states are reverse 
in  $^8\LLi$.
As for the splittings in $^{10}\LBe$, 
our result is similar to the SM predictions in Refs.~\cite{Millener:2012zz,Umeya-10Be}. 
On the other hand,
the four-body OCM cluster model calculation \cite{Hiyama:2012sq} gives quite small values in  $^{10}\LBe$
as 0.08 MeV and 0.05 MeV for the $2^-$-$1^-$ and $3^-$-$2^-$ splittings, which are inconsistent with our result
and SM predictions.

In $^{8}\LLi$, $^{10}\LBe$, $^{12}\LB$, $^{12}\LC$, and $\LO$, 
the $1^-$-$0^-$ splittings for core $I^\pi=1/2^-$ states
are remarkably small because significant negative contributions of the tensor ($T$)  term cancels the $\Delta_\sigma$ contribution.
As pointed out by Millener, 
this cancellation is essential to describe the small $1^-$-$0^-$ splitting observed in $\LO$.
Indeed, the Millener's SM calculation predicts the small $1^-$-$0^-$ splittings in $^{8}\LLi$, $^{10}\LBe$, $^{12}\LB$, and $^{12}\LC$,
and our result is consistent with it.
The $1^-$-$0^-$ splitting is experimentally known only for $\LO$ but not for other $\LZ$.
In both the present and Millener's SM calculations,
the value for $\LO$  is used as an input data in phenomenological tuning of the spin-dependent $\La N$ interactions.
In order to check validity of the tensor term of the spin-dependent $\La N$ interactions,  
experimental data for other systems are requested.

Let us turn to the $2^-$-$1^-$ splitting in $^{12}\LB(^{12}\LC)$  
for the excited core, $I^\pi=3/2^-_2$ state.
The nuclear intrinsic-spin and orbital angular momentum coupling in 
the $^{11}\CB(3/2^-_2)(^{11}\CC(3/2^-_2))$ state is quite different from that in the ground state, $I^\pi=3/2^-_1$ 
(cf. Table \ref{tab:mu}). 
The $I^\pi=3/2^-_2$ state dominantly contains the nuclear orbital angular momentum $L=2$ 
excitation coupled with a $p$-orbit valence nucleon
as $[p_{3/2,1/2}\otimes L=2]_{J=3/2}$ as indicated by 
the large $\langle \bvec{L}^2\rangle$ in Table \ref{tab:mu}. 
Because of the significant coupling with $L=2$,  the nuclear intrinsic-spin of the $p_{3/2}$ valence
nucleon is not aligned to the $I$ direction, and therefore, gives a relatively small $\Delta_\sigma$ contribution 
to the $2^-$-$1^-$ splitting in $^{12}\LB(^{12}\LC)$. On the other hand, the $T$ and $S_\La$ terms give 
negative contributions. As a result, 
the AMD (AMD') calculation give the negative values $-0.045(-0.071)$ and $-0.047(-0.073)$ MeV of the total $2^-$-$1^-$ splittings
namely, the reverse ordering for the core states $^{11}\CB(3/2^-_2)$ and $^{11}\CC(3/2^-_2)$, respectively
(cf. Tables \ref{tab:splitting-orig} and \ref{tab:nuclear-int-dep}).
The experimental value of this splitting has not been  determined yet. 
The $1^-$ state at 6.050 MeV in $^{12}\LC$ has been determined by the $\gamma$-ray measurement
\cite{Hosomi:2015fma}, 
whereas the $1^-$ and $2^-$ states in  $^{12}\LB$ are not separated but both are included in the peak observed at 
$5.92 \pm 0.13$ MeV in the $(e,e'K^+)$ reaction experiment
\cite{Tang:2014atx}. 
Provided that the Coulomb shift between mirror states in $^{12}\LC$-$^{12}\LB$ is the same as that in $^{11}\CC$-$^{11}\CB$, 
the $2^-$-$1^-$ splitting for $I^\pi=3/2^-_2$ is estimated to be $-0.215$ MeV. The reverse ordering is consistent 
with the AMD prediction, but quantitatively the agreement is not so good.
The negative splitting (reverse ordering) is also predicted by the SM calculation (theoretical value is $-0.122$ MeV) 
\cite{Motoba:2017lkk}.
More detailed experimental spectra are demanded to determine the splitting energy. 
 
\subsubsection{Case of even-even $Z=N$ core nuclei: $5/2^+$-$3/2^+$ splitting in 
$^9\LBe$ and $^{13}\LC$}

For $^9\LBe$ and $^{13}\LC$, we discuss 
spin-dependent contributions in the $5/2^+$-$3/2^+$ splittings 
for the excited core states,  $^8\CBe(2^+_1)$ and $^{12}\CC(2^+_1)$. 
The splittings are generally small because 
$n\alpha$-cluster structures are favored and  
the nuclear intrinsic-spin is almost saturated.
In $^9\LBe$, since the core state $^8\CBe(2^+_1)$ 
has the ideal $2\alpha$-cluster structure with 
the relatively $L=2$ wave,
the $\Delta_\sigma$ and $T$ contributions almost vanish and 
only the $S_\La$ term contributes to the splitting. 
It means that the $5/2^+$-$3/2^+$ splitting in $^9\LBe$ can be 
a sensitive probe to test the $S_\La$ term of the $\La N$ interactions. 
A tiny splitting $-0.023$ MeV is obtained in the AMD calculation. 
In the present calculation, the strength of the $S_\La$ term is not modified.
If the strength is tuned to fit the observed value $-0.043$,  
a slightly stronger $S_\La$ term by a factor of $\sim 2$ is favored.
This modification of the $S_\La$ term gives only minor effects to the splittings in other systems.

In $^{13}\LC$, the core state 
$^{12}\CC(2^+_1)$ contains the slight $S\ne 0$ component
because of $3\alpha$-cluster breaking
as can be seen in non-zero expectation value
$\langle\bvec{S}^2 \rangle\ne 0$ in Table \ref{tab:mu}.
The $S\ne 0$ component from the cluster breaking gives  
non-negligible $\Delta_\sigma$ and $T$ contributions, which
cancel the negative $S_\La$ contribution in the  
$5/2^+$-$3/2^+$ splitting (see Table \ref{tab:nuclear-int-dep}). 
Consequently, the predicted $5/2^+$-$3/2^+$ splitting in $^{13}\LC$ 
is a small positive value, 
0.064 MeV in the AMD and 0.026 MeV in the AMD' result. The difference 
between the AMD and AMD' results 
originates in the nuclear spin-orbit interaction dependence of the 
cluster breaking in the core state. 
If the twice stronger $S_\La$ interaction adjusted to the 
$5/2^+$-$3/2^+$ splitting in $^9\LBe$ is adopted,
further cancellation occurs in $^{13}\LC$.

The $\La N$ spin-orbit splittings in  $^{9}\LBe$ and $^{13}\LC$ have been investigated by the
$2\alpha$- and $3\alpha$-cluster OCM calculations \cite{Hiyama:2000jd}. 
In the OCM cluster model calculations, 
the nuclear intrinsic-spin completely vanishes and 
only the $S_\La$ term contributes to the $5/2^+$-$3/2^+$ splittings 
because $\alpha$ clusters are assumed.
The cluster OCM calculations with the NSC97f predicted the negative $5/2^+$-$3/2^+$ splittings as 
$-0.16$ MeV in $^{9}\LBe$ and $-0.29$ MeV in $^{13}\LC$. 
Compared with the later observed value $-0.043$ MeV in $^{9}\LBe$, 
the prediction suggests that 
the $S_\La$ term of the NSC97f interaction may be too strong.

\begin{figure}[!htb]
\begin{center}
\includegraphics[width=7.0cm]{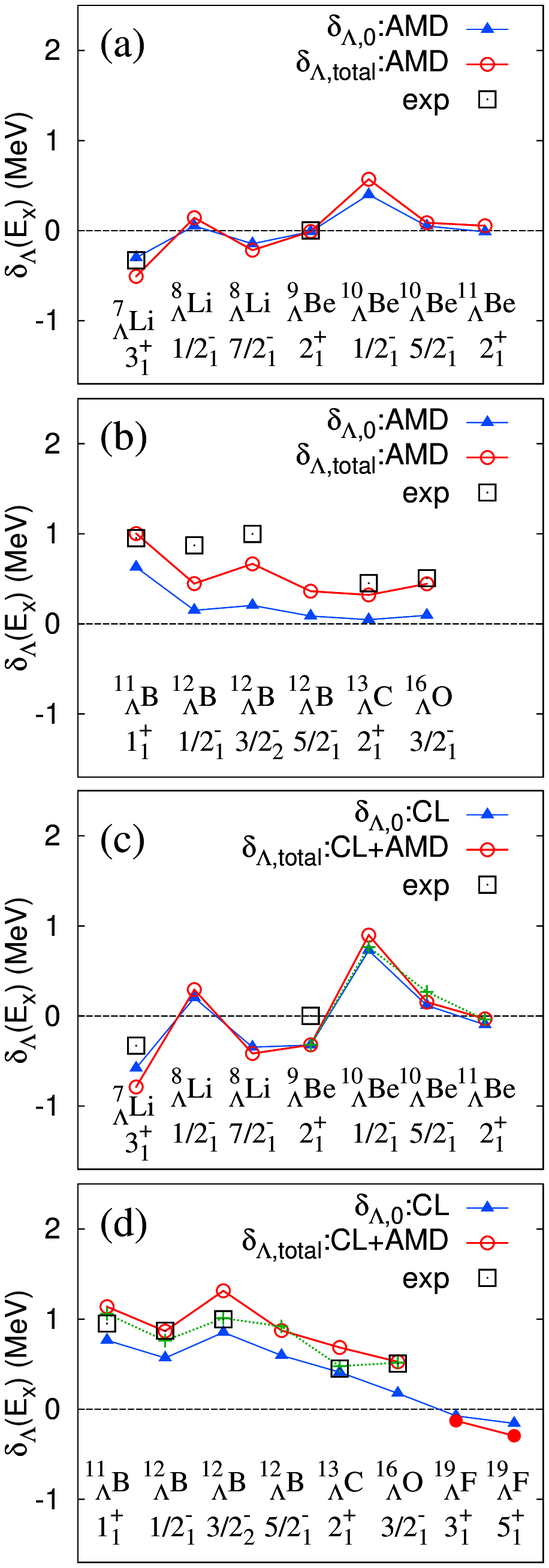} 	
\end{center}
  \caption{(color online) 
Spin-averaged excitation energy shift $\delta_\Lambda(\bar{E}_x)$ in $\LZ$ for core $\CZ(I^\pi)$ states. 
(a)(b) The total energy shift ($\delta_\La(\bar{E}_x)=\delta_{\La}(\bar{E}_x)+\delta_{\La,0}(\bar{E}_x)$)
and the $V_0$ contribution ($\delta_{\La,0}(\bar{E}_x)$) obtained by the AMD calculation. 
(c)(d) The total energy shift 
obtained by the CL+AMD calculation, and 
the $V_0$ contribution obtained by the CL calculation. 
In (c) and (d), 
the total energy shift obtained by the CL+AMD' calculation is also shown by green cross points. For 
$\LF$, the total energy shift and $V_0$ contribution 
obtained by the CL calculation are plotted by red filled circles and blue filled triangles, respectively.
Information of the experimental data is explained in the caption of Table \ref{tab:excitation-energy-shift}.
\label{fig:ex-s}}
\end{figure}

\begin{table*}[ht]
\caption{The spin-averaged excitation energy shifts ($\delta_{\La}(\bar{E}_x)$), 
and $S_N$ and  $V_0$ contributions ($\delta_{\La,S_N}(\bar{E}_x)$ and $\delta_{\La,S_0}(\bar{E}_x)$)
in $\LZ$ for core $\CZ(I^\pi)$ states.
The AMD result of the $S_N$ contribution, the AMD and CL results of the $V_0$ contribution, and 
the AMD and CL+AMD results of the total shifts are listed. 
The expectation values of the $V_{S_N}$ term obtained by the AMD calculation are also shown. 
For $\LF$, the CL and  CL$^{4\alpha}$+CL results are shown. 
The experimental values are from Refs.~\cite{Tamura:2010zz,Chrien:1990ag,Tanida:2000zs,Ukai:2006zp,Akikawa:2002tm,
Ajimura:2001na,Miura:2005mh,Ukai:2008aa,Ma:2010zzb,Kohri:2001nc,Tang:2014atx,Hosomi:2015fma}
except for $^{12}\LC(I^\pi=1/2^)$,
$^{12}\LC(I^\pi=3/2^-_2)$, and 
$^{13}\LC(I^\pi=0^+)$. For $^{12}\LC(I^\pi=1/2^-)$ and $^{13}\LC(I^\pi=2^+)$, the excitation energy shifts for the 
 $^{12}\LC(1^-)$ and  $^{13}\LC(3/2^+)$ are shown. $^{*}$For $^{12}\LC(I^\pi=3/2^-_2)$, 
the averaged value of the observed excitation energies of $^{12}\LC(1^-) $\cite{Hosomi:2015fma} and  $^{12}\LB(2^-)$ \cite{Tang:2014atx} 
is deduced by assuming the same Coulomb shift between mirror states in  $^{12}\LC$-$^{12}\LB$ as that in 
$^{11}\CC$-$^{11}\CB$. Energies are in MeV.
\label{tab:excitation-energy-shift}
}
\begin{center}
\begin{tabular}{cc|cc|cc|cc|cccccccc}
\hline			
$\LZ$ & ($I^\pi$) & $\langle  V_{S_N}\rangle$ & $\delta_{\La,S_N}(\bar{E}_x)$ & \multicolumn{2}{c|}{$\delta_{\La,0}(\bar{E}_x)$} &
\multicolumn{2}{c|}{$\delta_{\La}(\bar{E}_x)$} & $\delta_{\La}(\bar{E}_x)$  \\
 &    &AMD &  AMD & AMD & CL & AMD &  CL+AMD & exp \\
$^7\LLi$	&$	(3^+)	$&$	-0.22 	$&$	-0.21 	$&$	-0.30 	$&$	-0.58 	$&$	-0.51 	$&$	-0.79 	$&$	-0.329 	$\\
$^7\LLi$	&$	(1^+_\textrm{gs})	$&$	-0.01 	$&$	-	$&$	-	$&$	-	$&$	-	$&$	-	$&$	-	$\\
&&&&&&&&\\
$^8\LLi$	&$	(7/2^-)	$&$	-0.15 	$&$	-0.07 	$&$	-0.15 	$&$	-0.34 	$&$	-0.22 	$&$	-0.42 	$&$	-	$\\
$^8\LLi$	&$	(3/2^-_\textrm{gs})	$&$	-0.08 	$&$	-	$&$	-	$&$	-	$&$	-	$&$	-	$&$	-	$\\
$^8\LLi$	&$	(1/2^-)	$&$	0.01 	$&$	0.09 	$&$	0.05 	$&$	0.20 	$&$	0.14 	$&$	0.29 	$&$	-	$\\
&&&&&&&&\\
$^9\LBe$	&$	(2^+)	$&$	-0.03 	$&$	0.00 	$&$	-0.01 	$&$	-0.32 	$&$	-0.01 	$&$	-0.32 	$&$	0.002 	$\\
$^9\LBe$	&$	(0^+_\textrm{gs})	$&$	-0.03 	$&$	-	$&$	-	$&$	-	$&$	-	$&$	-	$&$	-	$\\
&&&&&&&&\\
$^{10}\LBe$	&$	(5/2^-)	$&$	-0.11 	$&$	0.04 	$&$	0.05 	$&$	0.10 	$&$	0.09 	$&$	0.14 	$&$	-	$\\
$^{10}\LBe$	&$	(3/2^-_\textrm{gs})	$&$	-0.15 	$&$	-	$&$	-	$&$	-	$&$	-	$&$	-	$&$	-	$\\
$^{10}\LBe$	&$	(1/2^-)	$&$	0.03 	$&$	0.17 	$&$	0.40 	$&$	0.72 	$&$	0.57 	$&$	0.89 	$&$	-	$\\
&&&&&&&&\\
$^{11}\LBe$	&$	(2^+)	$&$	-0.41 	$&$	0.07 	$&$	-0.01 	$&$	-0.10 	$&$	0.05 	$&$	-0.03 	$&$	-	$\\
$^{11}\LBe$	&$	(0^+_\textrm{gs})	$&$	-0.47 	$&$	-	$&$	-	$&$	-	$&$	-	$&$	-	$&$	-	$\\
&&&&&&&&\\
$^{11}\LB$	&$	(3^+_\textrm{gs})	$&$	-0.44 	$&$	-	$&$	-	$&$	-	$&$	-	$&$	-	$&$	-	$\\
$^{11}\LB$	&$	(1^+)	$&$	-0.07 	$&$	0.37 	$&$	0.63 	$&$	0.77 	$&$	1.00 	$&$	1.14 	$&$	0.951 	$\\
&&&&&&&&\\
$^{12}\LC$	&$	(5/2^-)	$&$	-0.30 	$&$	0.27 	$&$	0.08 	$&$	0.62 	$&$	0.36 	$&$	0.90 	$&$	-	$\\
$^{12}\LC$	&$	(3/2^-_\textrm{gs})	$&$	-0.57 	$&$	-	$&$	-	$&$	-	$&$	-	$&$	-	$&$	-	$\\
$^{12}\LC$	&$	(1/2^-)	$&$	-0.28 	$&$	0.29 	$&$	0.15 	$&$	0.59 	$&$	0.45 	$&$	0.89 	$&$	0.732 (1^-)	$\\
$^{12}\LC$	&$	(3/2^-_2	)$&$	-0.12 	$&$	0.46 	$&$	0.20 	$&$	0.89 	$&$	0.66 	$&$	1.35 	$&$	1.01^{*}	$\\
&&&&&&&&\\
$^{13}\LC$	&$	(2^+)	$&$	-0.40 	$&$	0.27 	$&$	0.05 	$&$	0.41 	$&$	0.32 	$&$	0.69 	$&$	0.451 (3/2^+)	$\\
$^{13}\LC$	&$	(0^+_\textrm{gs})	$&$	-0.67 	$&$	-	$&$	-	$&$	-	$&$	-	$&$	-	$&$	-	$\\
&&&&&&&&\\
$\LO$	&$	(3/2^-)	$&$	0.11 	$&$	0.35 	$&$	0.09 	$&$	0.18 	$&$	0.44 	$&$	0.53 	$&$	0.507 	$\\
$\LO$	&$	(1/2^-_\textrm{gs})	$&$	-0.24 	$&$	-	$&$	-	$&$	-	$&$	-	$&$	-	$&$	-	$\\
&&&&&&&&\\ 
&    & CL&  CL &CL & CL$^{4\alpha}$ & CL &   CL$^{4\alpha}$ +CL & \\																					
$\LF$	&$	(5^+)	$&$	-0.19 	$&$	-0.14 	$&$	-0.13 	$&$	-0.16 	$&$	-0.27 	$&$	-0.29 	$&$	-	$\\
$\LF$	&$	(3^+)	$&$	-0.11 	$&$	-0.05 	$&$	-0.04 	$&$	-0.07 	$&$	-0.10 	$&$	-0.13 	$&$	-	$\\
$\LF$	&$	(1^+_\textrm{gs})	$&$	-0.05 	$&$	-	$&$	-	$&$	-	$&$	-	$&$	-	$&$	-	$\\
\hline
\end{tabular}
\end{center}
\end{table*}	

\subsection{$S_N$ contributions to energy spectra and $B_\La$}

\subsubsection{Excitation energy shifts} 

In the present perturbative treatment of the spin-dependent part $V_1$ 
of the $\La N$ interactions, 
the $S_N$ term gives no contribution to the spin-doublet splitting but 
contributes to the spin-averaged excitation energy shift
$\delta_\La (\bar{E}_x(I^\pi))$. 
The spin-averaged excitation energy shift  
originates in the nuclear structure difference between the 
ground and excited states, and is given by two contributions, 
the $V_0$ contribution ($\delta_{\La,0}(\bar{E}_x)$)
and the $S_N$ contribution ($\delta_{\La,S_N}(\bar{E}_x)$), 
as explained in Eq.~\eqref{eq:ex-shift}.
The $V_0$ contribution reflects mainly the 
difference in the core nuclear size, whereas 
the $S_N$ contribution is sensitive to that 
in the nuclear intrinsic-spin and orbital angular momentum 
configurations.

Table \ref{tab:excitation-energy-shift} shows the $V_0$ and $S_N$ contributions in $\delta_\La (\bar{E}_x)$
as well as total values obtained by AMD and CL+AMD calculations together with the experimental data. 
In both the calculations, the $S_N$ contributions are calculated with the AMD except for $\LF$
as explained previously. 
The expectation values 
$\langle V_{S_N} \rangle$ calculated with the AMD are also shown in the table.
In Fig.~\ref{fig:ex-s}, 
the $V_0$ contributions and the total excitation energy shifts are shown
compared with observed values of $\delta_{\La}(\bar{E}_x)$.
As shown in the table and figure, the CL calculation tends to give larger $V_0$ contributions than the AMD calculation because it 
gives larger size difference between the ground and excited states. 
The $S_N$ contributions 
are relatively minor in light-mass nuclei, whereas they are comparable or even larger than the $V_0$ contributions 
in heavy-mass nuclei. 
Both the AMD and CL+AMD calculations qualitatively describe
systematic behavior of  the experimental 
$\delta_{\La}(\bar{E}_x)$ in $p$-shell $\LZ$.
Quantitatively, 
the CL+AMD (AMD) calculation 
more or less overestimates (underestimates) the observed data
in the $A\ge 11$ region. 

In Figs.~\ref{fig:ex-s}(c) and (d), 
we also show the CL+AMD' result of 
 $\delta_{\La}(\bar{E}_x)$ to see the $NN$ spin-orbit interaction dependence.
Note that the difference between the CL+AMD and CL+AMD' 
calculations is the difference in the $S_N$ contribution
between the AMD and AMD' calculations.
In most states 
except for $^{12}\LB(I^\pi=3/2^-_2)$ and $^{13}\LC(I^\pi=2^+)$, 
difference between two calculations 
is rather small indicating that the dependence in
the excitation energy shift is minor. 
However, significant difference
is found in $^{12}\LB(I^\pi=3/2^-_2)$ and $^{13}\LC(I^\pi=2^+)$, 
in which nuclear intrinsic-spin configurations are sensitive to the $NN$ spin-orbit interaction 
as seen in Table \ref{tab:excitation-energy-shift}. In these states, the $S_N$ contribution is smaller
in the AMD' calculation because of the less cluster breaking  than the AMD calculation. 

Let us discuss the $^{16}\textrm{O}$ core vibration effect to the $V_0$ contribution in $\LO$ and $\LF$, 
which are taken into account in the CL calculation of  $\LO$ and the CL-$4\alpha$ calculation of $\LF$. 
For $\LF$, the 
CL-$4\alpha$ result 
is compared with the CL one, and for $\LO$, the AMD result is compared with the CL+AMD one
in Table \ref{tab:excitation-energy-shift}.
In both systems, the $4\alpha$ vibration gives only minor effect in the total excitation energy shifts. 

\subsubsection{$S_N$ contributions to binding energies $\bar{B}_\La$}

The $S_N$ term of the $\La N$ interactions also contributes to the $\La$ binding energies $\bar{B}_\La$. 
In particular, spin-orbit favored states in core nuclei gain much potential energy because the  $0s$-orbit $\La$ 
enhances the single-nucleon (mean) 
spin-orbit potential through the $\bvec{l}\cdot \bvec{s}_N$ term in $V_{S_N}$. Its expectation value 
in the ground state is nothing but the $S_N$ contribution to the $\La$ binding energies as 
$\bar{B}_{\La,S_N}=-\langle V_{S_N} \rangle$. 
As expected, 
the $S_N$ term gives non-negligible contributions to $\bar{B}_\La$ for 
the spin-orbit favored ground states such as $^{11}\LBe(I^\pi=0^+)$, $^{11}\LB(I^\pi=3^+)$, $^{12}\LB(I^\pi=3/2^-_\textrm{gs})$, $^{12}\LC(I^\pi=3/2^-_\textrm{gs})$, and $^{13}\LC(I^\pi=0^+)$, whereas 
it gives  minor contributions to well clustered states in light-mass $\LZ$  (see Table \ref{tab:excitation-energy-shift}).
There remains ambiguity in the result
because the $S_N$ contribution depends on the 
$NN$ spin-orbit interaction.
 For 
$^{11}\LBe(I^\pi=0^+)$, $^{11}\LB(I^\pi=3^+)$, $^{12}\LB(I^\pi=3/2^-_\textrm{gs})$, $^{12}\LC(I^\pi=3/2^-_\textrm{gs})$, 
and  $^{13}\LC(I^\pi=0^+)$, 
the AMD'  result of $\langle V_{S_N} \rangle$
is about half of the AMD result as 
$\langle V_{S_N} \rangle=-0.27$, $-0.36$, $-0.38$, $-0.38$ and $-0.30$ in MeV, respectively.
Another ambiguity comes from 
the $V_{S_N}$ term of the $\La N$ interactions, which has not been checked yet in the 
present work.

\begin{figure}[!htb]
\begin{center}
\includegraphics[width=8cm]{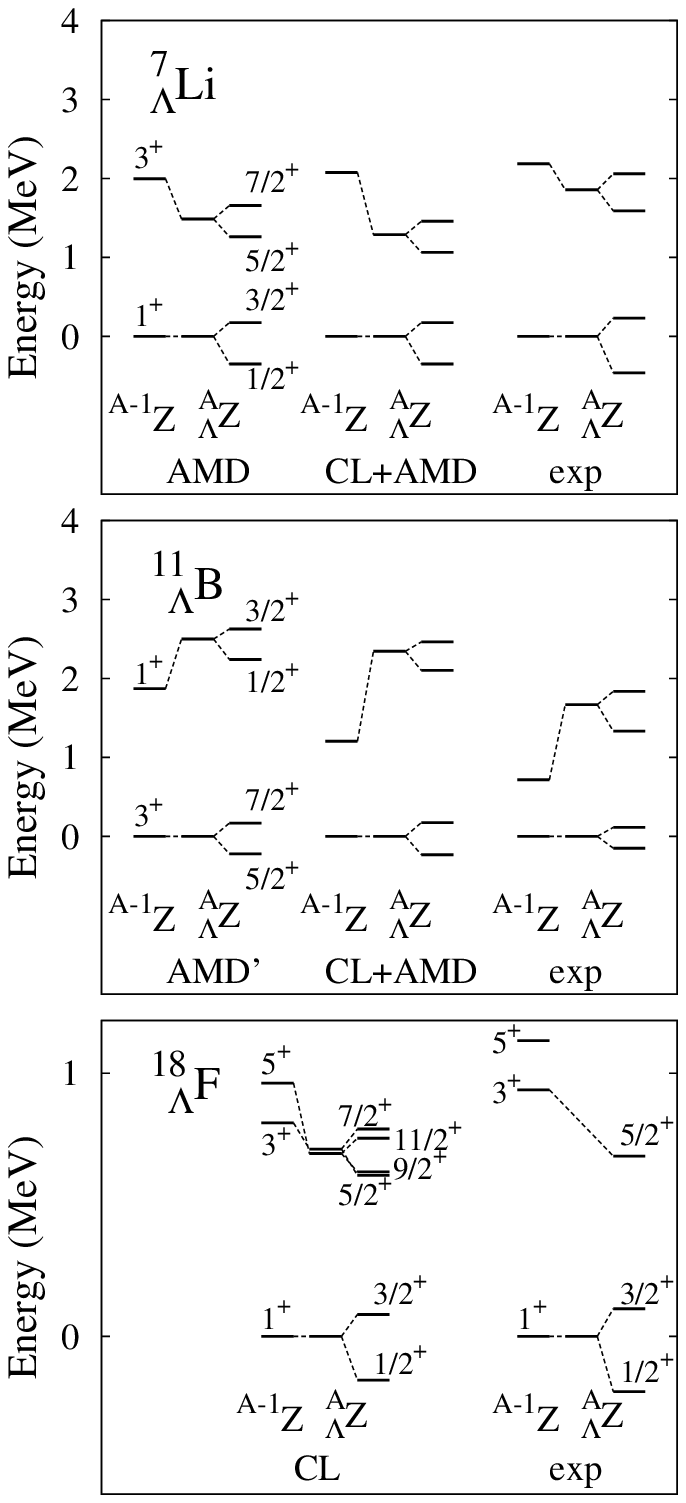} 	
\end{center}
  \caption{Energy spectra in $\LZ$
for $^7\LLi$, $^{11}\LB$, and $\LF$ together with those in 
$\CZ$. 
The spectra in $\CZ$, the spin-averaged spectra in $\LZ$, and spectra in $\LZ$
are shown in the left, middle, and right columns, respectively. 
Experimental data are from Refs.~\cite{Tamura:2010zz,Tanida:2000zs,Ukai:2006zp,Miura:2005mh,Ma:2010zzb,Yang:2017lay}.
\label{fig:spe1}}
\end{figure}

\begin{figure}[!htb]
\begin{center}
\includegraphics[width=8cm]{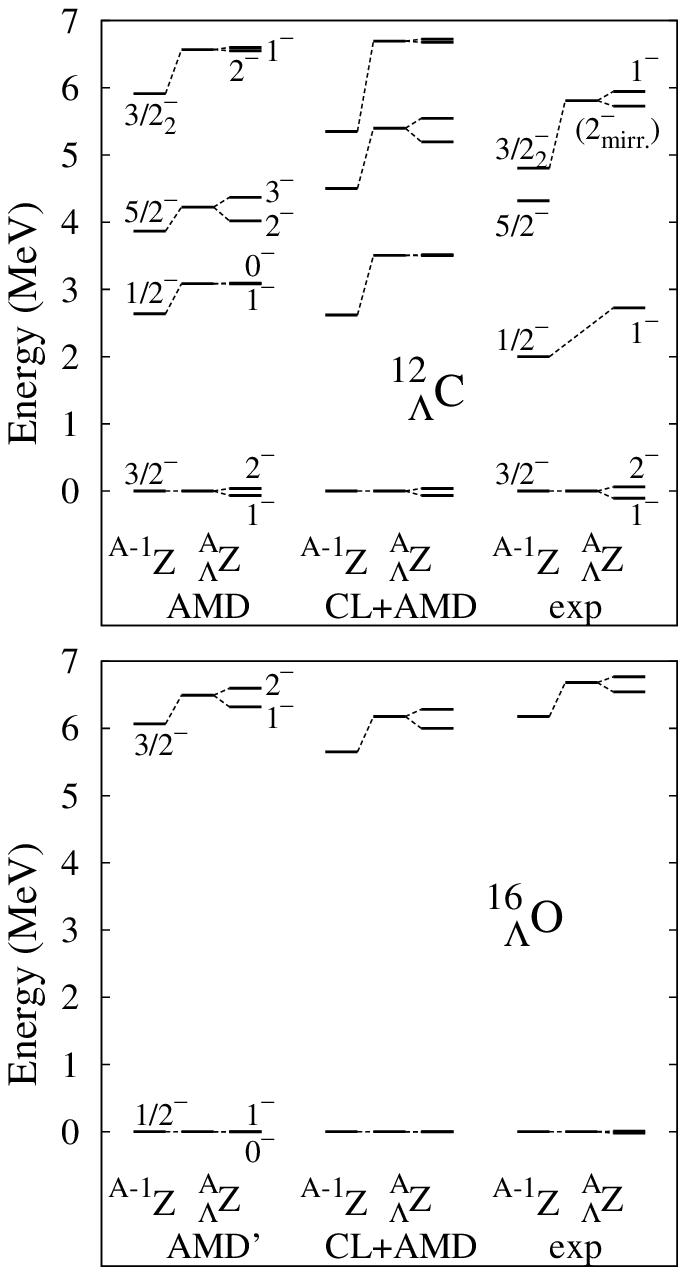} 	
\end{center}
  \caption{Same as Fig.~\ref{fig:spe2} but for 
 $^{12}\LC$ and  $\LO$. Experimental data are from Refs.~\cite{Tamura:2010zz,Hosomi:2015fma,Ukai:2008aa}.
As for the $^{12}\LC(J^\pi=2^-)$ state for the core $^{11}\CC(3/2^-_2)$ state, 
the observed energy of the mirror state in $^{12}\LB$ \cite{Tang:2014atx} is plotted assuming the
Coulomb shift between the mirror states in $^{12}\LC$-$^{12}\LB$ is the same as that in $^{11}\CC$-$^{11}\CB$.
\label{fig:spe2}}
\end{figure}

\subsection{Energy spectra}

Energy spectra in $^{7}\LLi$, $^{11}\LB$, $^{12}\LC$, $\LO$, and $\LF$ are shown in Figs.~\ref{fig:spe1} and ~\ref{fig:spe2}.
For $^{7}\LLi$ and $^{12}\LC$, spectra calculated with the AMD and CL+AMD are shown, and for
$\LF$ those with the CL calculation are shown. 
For $^{11}\LB$ and $\LO$, the spectra obtained with the CL+AMD and AMD'
are shown.
In the figures, the original energy spectra in $\CZ$, spin-averaged energy spectra in $\LZ$, 
and spectra in $\LZ$ are shown in the left, middle, and right columns, respectively.

\subsubsection{Energy spectra of $^{7}\LLi$ and $^{11}\LB$}

In the spin-averaged energy spectra in $^7\LLi$ and $^{11}\LB$, 
the excitation energy of $^{6}\CLi(3^+)$ is shifted downward in $^{7}\LLi(3^+)$, whereas 
that of $^{10}\CB(1^+)$ is shifted upward in $^{11}\CB(1^+)$ because of the $V_0$ and $S_N$ contributions 
(cf. Table \ref{tab:excitation-energy-shift}). 
In the energy spectra in $^7\LLi$ and $^{11}\LB$, the significant spin-doublet splitting occurs mainly 
because of the $\Delta_\sigma$ term contributed by $S=1$ $pn$ pairs 
around the  $\alpha$- and $2\alpha$-cluster structures, respectively. 

The present result of excitation energy shifts and spin-doublet splitting energies in  $^7\LLi$ and $^{11}\LB$
are  qualitatively consistent with the experimental spectra. 
Strictly speaking, however, 
the agreement with the energy spectra is not perfect. For example, 
the excitation energies of $^7\LLi(7/2^+)$ and $^7\LLi(5/2^+)$ are underestimated, 
in particular, by the CL calculation. 
The reason may be that the $s$-orbit $\La$ approximation is not enough for the core nucleus $^6\CLi$ 
having the remarkable $\alpha+d$ cluster structure, 
and may overestimate the energies of $^7\LLi(3/2^+)$ and  $^7\LLi(1/2^+)$ for the core $^6\CLi(1^+)$ state.
In order to discuss detailed energy spectra in $^7\LLi$,  
more precise calculations with microscopic three-body or four-body cluster models are needed
needed as has been tried in Refs.~\cite{Hiyama:1999me,Hiyama:2006xv}. 

\subsubsection{Energy spectra of $^{12}\LC$}
Available data of energy spectra in $^{12}\LC$ are reasonably reproduced by the calculations. 
The excitation energies for $^{11}\CC(1/2^-)$,  $^{11}\CC(5/2^-)$, and  $^{11}\CC(3/2^-_2)$ 
are significantly raised because of the stronger binding energy between $\Lambda$ and 
$^{11}\CC(3/2^-_{gs})$  (cf. Table \ref{tab:BE-radii}).
The spin-doublet splittings for the core $^{11}\CC(3/2^-_\textrm{gs})$ and $^{11}\CC(5/2^-)$
are moderate because, as a leading oder, one valence nucleon spin contributes to the splittings. 
The spin-doublet splittings have not been measured yet except for the $2^-_1$-$1^-_1$ splitting in 
$^{12}\LC$. 
For the core excited state, $^{11}\CC(3/2^-_2)$, the negative splitting energy, i.e., the reverse ordering of the spin-doublet states
is predicted. The present prediction is supported by 
the experimental  excitation energy of the mirror $2^-$ state in $^{12}\LB$ measured by the 
production cross section analysis \cite{Tang:2014atx}. 

\subsubsection{Energy spectra of $\LO$}
The experimental spectra in $\LO$ are reproduced well by the calculations.
The spin-averaged excitation energy for the core $\CO(3/2^-)$ state is shifted upward by the $\La$
mainly because of the $S_N$ contribution. The $V_0$ contribution to the excitation energy shift is minor. 
The AMD' and CL+AMD results are similar 
indicating that the vibration effect in the $^{16}\textrm{O}$ core is minor
because the $\CO$ structure is rather robust
differently from fragile cluster structures of  light-mass $p$-shell nuclei.
The spin-doublet splittings for the core states, $\CO(1/2^-)$ and $\CO(3/2^-)$, are 
tiny and moderate, respectively, reflecting the nuclear spin configurations in the core nucleus. 

\subsubsection{Energy spectra of $\LF$}
The recently observed 
0.315 MeV and 0.895 MeV $\gamma$-rays are assigned to the  $M1$ $3/2^+\to 1/2^+$ and  $E2$
$5/2^+\to 1/2^+$ transitions, respectively \cite{Yang:2017lay}.
The calculated value 0.25 MeV
of the  $3/2^+$-$1/2^+$ splitting  is in reasonable agreement with the 
observed value 0.315 MeV. The $3/2^+$-$1/2^+$ splitting is contributed mainly by the $\Delta_\sigma$ term,
which reflects the dominant nuclear intrinsic-spin $S=1$ component of a $pn$ pair in the $L=0$ wave.
For the $7/2^+$ and $5/2^+$ states, a smaller splitting 0.18 MeV than the  $3/2^+$-$1/2^+$ splitting is predicted
because of the cancellation of the $\Delta_\sigma$ contribution by the $T$ and $L_\La$ contributions
as discussed previously.

Let us discuss the detail of the excitation energy $E_x(5/2^+)$ in $\LF$. The experimental 
energy shift $-0.042$ MeV of $\LF(5/2^+)$ is reduced from the observed $E_x(5/2^+)=0.895$ MeV in $\LF$
and $E_x(3^+)$=0.937 MeV in $\CF$.
The calculated energy shift is $-0.03$ ($-0.07$) MeV in the CL 
(CL$^{4\alpha}$+CL) calculations. The result reasonably agrees with
the experimental data. 
In the CL calculation, 
the $S_N$ and $V_0$ contributions are 
$-0.05$ MeV and $-0.04$ MeV, respectively. In addition, 
the spin-doublet splitting energy gives positive contribution of $+0.07$ MeV to the excitation energy shift
because it causes larger energy gain in the ground $\LF(1/2^+)$ state than the $\LF(5/2^+)$ state.
For more detailed discussion, 
the experimental measurement of the excitation energy for the spin-doublet partner $\LF(7/2^+)$ is highly requested.
 
\section{Summary} \label{sec:summary}
Energy spectra of $0s$-orbit $\La$ states in $p$-shell $\Lambda$ hypernuclei ($\LZ$) and 
those in $\LF$ were studied 
with the AMD+VAP and microscopic cluster model
using the $\La NG$ interactions. 
The spin-dependent terms of the 
ESC08a $\La N G$ interaction were tested 
in comparison of the calculated energy spectra with the observed ones. 
A modification of the spin-dependence of the ESC08a $\La NG$ interaction 
was proposed by phenomenological tuning of the spin-spin $(\sigma_\La\cdot \sigma_N)$ and tensor ($S_{12}$) 
terms to adjust available data of energy spectra in $p$-shell $\LZ$.

The spin-dependent 
contributions of the $\La N$ interactions to spin-doublet splittings were discussed.  
In the case of odd-odd, even-odd, and odd-even core nuclei, 
the $\Delta_\sigma$ contribution is usually dominant, whereas the $T$ and $S_\La$ contributions
are relatively minor in most cases. There are some exceptions such as $I^\pi=1/2^-$ states in
even-odd and odd-even core nuclei, in which the significant tensor contribution cancels the $\Delta_\sigma$ contribution. 
In $^{13}\LC$,  the cluster breaking component gives non-negligible contributions to the splitting energy for 
the core $^{12}\CC(2^+)$ state through the $\Delta_\sigma$  and tensor terms of the $\La N$ interactions. 
The $V_0$ and $S_N$ contributions to the excitation energy shifts 
were also discussed. Calculated energy spectra as well as spin-averaged energy spectra in $\LZ$ were compared with 
experimental data. The calculations reasonably reproduce the observed spectra  in $p$-shell $\LZ$ and $\LF$. 

The extensive data of $p$-shell $\LZ$ observed by high-resolution $\gamma$-ray measurements 
are useful information to obtain comprehensive understanding of the energy spectra and structures of $\La$ hypernuclei.
Moreover, they are useful to test the effective $\La N$ interactions in hypernuclei.  In particular, the spin-doublet splitting 
is good probe to check the spin-dependence of the $\La N$ interactions. 
In the present systematic investigation of energy spectra in $\LZ$, we proposed the modified spin-dependent 
$\La N$ interactions which can reasonably reproduce the observed spin-doublet splitting energies in $p$-shell
$\LZ$. The present work may shed a light on spin-dependence of the effective $\La N$ interactions in $p$-shell 
$\LZ$ and enable us to predict spectra for unobserved excited states.
In order to explore such systematic investigations of $\La$ hypernuclei in a wide mass number region,
further $\gamma$-ray spectroscopic studies of $\La$ hypernuclei in heavier mass regions are requested.

In the present work, two models,  the AMD+VAP and microscopic cluster models, were adopted. 
The former is useful to treat  nuclear intrinsic-spin configurations in detail, whereas 
the latter is suitable to describe the dynamical inter-cluster motion.
In order to investigate energy spectra in 
$\LZ$ precisely,  further advanced frameworks that can describe details of  intrinsic-spin configurations as well as
dynamical structure change are needed. The HAMD method is one of the promising tools.  
The ambiguity in the effective $NN$ interactions is also a  remaining problem to be solved.

\appendix
\section{Comparison between the folding potential approximation and the microscopic calculation of $V_0$} \label{app:folding-pot}

In the present calculation, the $\Lambda$ wave functions are obtained 
by folding the spin-independent central term 
($V_0$) of the $\La N$ interactions. The nuclear density matrix in the exchange potential is approximated
with the density matrix expansion in the local density approximation  \cite{Negele:1975zz}.
We discuss here its validity of the approximation for energy spectra 
 in $^7\LLi$.

In Table \ref{tab:V0-approx}, 
the approximated $\La$-potential energy $\langle\phi_{\La,0} | U_0 (\rho_N^{I^\pi})|\phi_{\La,0} \rangle$ is
compared with the microscopically  calculated energy $\langle \Psi_{\LZ}(J^\pi)  | V_0|\Psi_{\LZ}(J^\pi)  \rangle$.
Here,  $\Psi_{\LZ}(J^\pi)= \left[\Psi_N(I^\pi_n)  \phi_{\La,0}\chi_\La \right ]_J$ is the microscopic $A$-body wave function 
for $\LZ$ and 
$\phi_{\La,0}(\rho_N^{I^\pi};r)$ is fixed to be that obtained by the folding potential model.
It should be noted that, in the folding potential model,  
the nuclear density $\rho_N^{I^\pi}$ is defined for intrinsic wave functions of $\CZ$ without the cm motion, and 
the $\La$ recoil effect is properly taken into account.
However,  the microscopic $A$-body wave function  $\Psi_{\LZ}$ contains the cm motion.
For consistency, 
we also perform the approximated and the microscopic calculations of the $\La$-potential energy
by using the nuclear density  $\rho_{N,\textrm{cm}}^{I^\pi}$ with the cm motion instead of $\rho_{N}^{I^\pi}$
without the cm motion. 
As shown in the table,  errors of the approximation are only $<7\%$ and $<2\%$ 
in the $\La$-potential energy calculated with $\rho_{N}^{I^\pi}$ and $\rho_{N,\textrm{cm}}^{I^\pi}$, 
respectively.  Moreover, the errors are almost state-independent and give only 
global shifts meaning that the approximation gives minor effect to energy spectra. 

\begin{table}[ht]
\caption{Comparison of the $\La$ potential energy in  $^7\LLi(I^\pi=1^+,3^+)$ 
between the folding potential model approximation  and
the microscopic calculation.
The energies calculated using the nuclear densities $\rho_{N}^{I^\pi}$
and  $\rho_{N,\textrm{cm}}^{I^\pi}$ (without and with the cm motion) are shown. 
The difference between approximated and microscopic calculations is also shown.
Energies are in MeV.
\label{tab:V0-approx}
}
\begin{center}
\begin{tabular}{cccccccccccccc}
\hline			
 &	& approx. & micro. & diff. \\
$^7\LLi(I^\pi=1^+)$ &$\rho_N$	&$	-11.81 	$&$	-11.04 	$&$	-0.77 	$\\
& $\rho_{N,\textrm{cm}}$&$	-11.10 	$&$	-10.92 	$&$	-0.18 	$\\
& & \\
$^7\LLi(I^\pi=3^+)$ & $\rho_N$&$	-12.90 	$&$	-12.06 	$&$	-0.84 	$\\
& $\rho_{N,\textrm{cm}}$&$	-12.16 	$&$	-11.99 	$&$	-0.17 	$\\

\hline			
\end{tabular}
\end{center}
\end{table}

\section{Core rearrangement effect to spin-doublet splitting energy} \label{app:rearrangements}
 In the present  perturbative treatment of the spin-dependent part ($V_1$) of the $\La N$ interactions, 
the nuclear spin rearrangement is ignored. 
Generally, its effects are expected to be minor because $V_1$ is 
relatively weak compared with the $NN$ interactions and also the spin-independent part  ($V_0$) of the 
$\La N$ interactions. 
Possible exception is the case that two energy levels with the same 
$J^\pi$ eventually exist close to each other.
In order to see nuclear spin rearrangement effects,  
we calculate the energy spectra of $^8\LLi$, $^{10}\LBe$, $^{12}\LC$, and $\LO$ with the AMD
by diagonalization of the full Hamiltonian 
including $V_0$ and $V_1$ terms of the $\La N$ interactions.
The spin-doublet splittings calculated with and without the rearrangement
are compared in Table \ref{tab:rearrangement}.
One can see that the rearrangement effect is minor in most of states.

\begin{table}[ht]
\caption{Spin-doublet splittings in $\LZ$ for core $\CZ(I^\pi)$ states calculated with and without  nuclear spin rearrangement
in the AMD calculation. 
\label{tab:rearrangement}
}
\begin{center}
\begin{tabular}{cccccccccccccc}
\hline
    $\LZ$    & ($I^\pi$) & $J^\pi_>$ & $J^\pi_<$  & \multicolumn{3}{c}{Splitting energy (MeV)} & \\ 
 &  &  &  &w/o 	&	with 		\\
$^8\LLi$	&$	(7/2^-)	$&$	4^-	$&$	3^-	$&$	0.22 	$&$	0.25 	$\\
&$	(5/2^-)	$&$	3^-	$&$	2^-	$&$	-0.11 	$&$	-0.10 	$\\
	&$	(3/2^-_\textrm{gs})	$&$	2^-	$&$	1^-	$&$	0.25 	$&$	0.30 	$\\
	&$	(1/2^-)	$&$	1^-	$&$	0^-	$&$	-0.03 	$&$	0.00 	$\\
& &&&& \\
$^{10}\LBe$	&$	(5/2^-)	$&$	3^-	$&$	2^-	$&$	0.20 	$&$	0.21 	$\\
	&$	(3/2^-_\textrm{gs})	$&$	2^-	$&$	1^-	$&$	0.16 	$&$	0.17 	$\\
	&$	(1/2^-)	$&$	1^-	$&$	0^-	$&$	0.00 	$&$	0.00 	$\\
& &&&& \\
$^{12}\LC$	&$	(5/2^-)	$&$	3^-	$&$	2^-	$&$	0.35 	$&$	0.36 	$\\
	&$	(3/2^-_\textrm{gs})	$&$	2^-	$&$	1^-	$&$	0.11 	$&$	0.12 	$\\
	&$	(1/2^-)	$&$	1^-	$&$	0^-	$&$	-0.01 	$&$	-0.01 	$\\
	&$	(3/2^-_2)	$&$	2^-	$&$	1^-	$&$	-0.05 	$&$	-0.05 	$\\
& &&&& \\
$\LO$	&$	(3/2^-)	$&$	2^-	$&$	1^-	$&$	0.28 	$&$	0.28 	$\\
	&$	(1/2^-_\textrm{gs})	$&$	1^-	$&$	0^-	$&$	0.01 	$&$	0.00 	$\\
\hline			
\end{tabular}
\end{center}
\end{table}

\begin{acknowledgments}
The computational calculations of this work were performed by using the
supercomputer in the Yukawa Institute for theoretical physics, Kyoto University. This work was supported by 
Grant-in-Aid for Scientific Research (C) (No. 26400270),
Grants-in-Aid for Young Scientists (B) (No. 15K17671), and 
Grant-in-Aid for JSPS Research Fellow (No. 16J05297) from Japan Society for the Promotion of Science(JSPS). 

\end{acknowledgments}


\begin{thebibliography}{9}
\bibitem{Hashimoto:2006aw} 
  O.~Hashimoto and H.~Tamura,
  Prog.\ Part.\ Nucl.\ Phys.\  {\bf 57}, 564 (2006).

\bibitem{Tamura:2010zz} 
  H.~Tamura,
  Prog.\ Theor.\ Phys.\ Suppl.\  {\bf 185}, 315 (2010).

\bibitem{Tamura:2013lwa} 
  H.~Tamura {\it et al.},
  Nucl.\ Phys.\ A {\bf 914}, 99 (2013).

\bibitem{Yang:2017lay} 
  S.~B.~Yang {\it et al.},
  JPS Conf.\ Proc.\  {\bf 17}, 012004 (2017).


\bibitem{Motoba:1984ri} 
  T.~Motoba, H.~Band\=o and K.~Ikeda,
  Prog.\ Theor.\ Phys.\  {\bf 70}, 189 (1983).
\bibitem{motoba85}
T. ~Motoba, H. Band\=o, K. Ikeda and T. Yamada,  Prog.\ Theor.\ Phys.\ Suppl. {\bf 81} 42 (1985). 

\bibitem{Yamada:1985qr} 
  T.~I.~Yamada, K.~Ikeda, H.~Bando and T.~Motoba,
  Prog.\ Theor.\ Phys.\  {\bf 73}, 397 (1985).

\bibitem{Yu:1986ip} 
  Y.~W.~Yu, T.~Motoba and H.~Bando,
  Prog.\ Theor.\ Phys.\  {\bf 76}, 861 (1986).


\bibitem{Hiyama:1996gv} 
  E.~Hiyama, M.~Kamimura, T.~Motoba, T.~Yamada and Y.~Yamamoto,
  Phys.\ Rev.\ C {\bf 53}, 2075 (1996).

\bibitem{Hiyama:1997ub} 
  E.~Hiyama, M.~Kamimura, T.~Motoba, T.~Yamada and Y.~Yamamoto,
  Prog.\ Theor.\ Phys.\  {\bf 97}, 881 (1997).

\bibitem{Hiyama:1999me} 
  E.~Hiyama, M.~Kamimura, K.~Miyazaki and T.~Motoba,
  Phys.\ Rev.\ C {\bf 59}, 2351 (1999).

\bibitem{Hiyama:2000jd} 
  E.~Hiyama, M.~Kamimura, T.~Motoba, T.~Yamada and Y.~Yamamoto,
  Phys.\ Rev.\ Lett.\  {\bf 85}, 270 (2000).

\bibitem{Hiyama:2002yj} 
  E.~Hiyama, M.~Kamimura, T.~Motoba, T.~Yamada and Y.~Yamamoto,
  Phys.\ Rev.\ C {\bf 66}, 024007 (2002).

\bibitem{Hiyama:2006xv} 
  E.~Hiyama, Y.~Yamamoto, T.~A.~Rijken and T.~Motoba,
  Phys.\ Rev.\ C {\bf 74}, 054312 (2006).

\bibitem{Hiyama:2010zzc} 
  E.~Hiyama, T.~Motoba, T.~A.~Rijken and Y.~Yamamoto,
  Prog.\ Theor.\ Phys.\ Suppl.\  {\bf 185}, 1 (2010).

\bibitem{Hiyama:2012sq} 
  E.~Hiyama and Y.~Yamamoto,
  Prog.\ Theor.\ Phys.\  {\bf 128}, 105 (2012).

\bibitem{Cravo:2002jv} 
  E.~Cravo, A.~C.~Fonseca and Y.~Koike,
  Phys.\ Rev.\ C {\bf 66}, 014001 (2002).

\bibitem{Suslov:2004ed} 
  V.~M.~Suslov, I.~Filikhin and B.~Vlahovic,
  J.\ Phys.\ G {\bf 30}, 513 (2004).

\bibitem{Mohammad:2009zza} 
  M.~Shoeb and Sonika,
  Phys.\ Rev.\ C {\bf 79}, 054321 (2009).


\bibitem{Zhang:2012zzg} 
  Y.~Zhang, E.~Hiyama and Y.~Yamamoto,
  Nucl.\ Phys.\ A {\bf 881}, 288 (2012).

\bibitem{Funaki:2014fba} 
  Y.~Funaki, T.~Yamada, E.~Hiyama, B.~Zhou and K.~Ikeda,
  Prog. Theor. Exp. Phys. {\bf 2014}, no. 11, 113D01 (2014).

\bibitem{Funaki:2017asz} 
  Y.~Funaki, M.~Isaka, E.~Hiyama, T.~Yamada and K.~Ikeda,
  Phys.\ Lett.\ B {\bf 773}, 336 (2017).

\bibitem{Gal:1971gb} 
  A.~Gal, J.~M.~Soper and R.~H.~Dalitz,
  Annals Phys.\  {\bf 63}, 53 (1971).

\bibitem{Gal:1972gd} 
  A.~Gal, J.~M.~Soper and R.~H.~Dalitz,
  Annals Phys.\  {\bf 72}, 445 (1972).
\bibitem{Gal:1978jt} 
  A.~Gal, J.~M.~Soper and R.~H.~Dalitz,
  Annals Phys.\  {\bf 113}, 79 (1978).

\bibitem{Millener:2008zz} 
  D.~J.~Millener,
  Nucl.\ Phys.\ A {\bf 804}, 84 (2008).

\bibitem{Millener:2010zz} 
  D.~J.~Millener,
  Nucl.\ Phys.\ A {\bf 835}, 11 (2010).

\bibitem{Millener:2012zz} 
  D.~J.~Millener,
  Nucl.\ Phys.\ A {\bf 881}, 298 (2012).


\bibitem{Guleria:2011kk} 
  N.~Guleria, S.~K.~Dhiman and R.~Shyam,
  Nucl.\ Phys.\ A {\bf 886}, 71 (2012).

\bibitem{Vidana:2001rm} 
  I.~Vidana, A.~Polls, A.~Ramos and H.-J.~Schulze,
  Phys.\ Rev.\ C {\bf 64}, 044301 (2001).

\bibitem{Zhou:2007zze} 
  X.~R.~Zhou, H.-J.~Schulze, H.~Sagawa, C.~X.~Wu and E.~G.~Zhao,
  Phys.\ Rev.\ C {\bf 76}, 034312 (2007).
\bibitem{Win:2008vw} 
  M.~T.~Win and K.~Hagino,
  Phys.\ Rev.\ C {\bf 78}, 054311 (2008).

\bibitem{Win:2010tq} 
  M.~T.~Win, K.~Hagino and T.~Koike,
  Phys.\ Rev.\ C {\bf 83}, 014301 (2011).

\bibitem{Lu:2011wy} 
  B.~N.~Lu, E.~G.~Zhao and S.~G.~Zhou,
  Phys.\ Rev.\ C {\bf 84}, 014328 (2011).
\bibitem{Mei:2014hya}
  H.~Mei, K.~Hagino, J.~M.~Yao and T.~Motoba,
  Phys.\ Rev.\ C {\bf 90},  064302 (2014).

\bibitem{Mei:2015pca} 
  H.~Mei, K.~Hagino, J.~M.~Yao and T.~Motoba,
  Phys.\ Rev.\ C {\bf 91}, 
  064305 (2015).

\bibitem{Mei:2016lce} 
  H.~Mei, K.~Hagino, J.~M.~Yao and T.~Motoba,
  Phys.\ Rev.\ C {\bf 93}, no. 4, 044307 (2016).

\bibitem{Schulze:2014oia} 
  H.-J.~Schulze and E.~Hiyama,
  Phys.\ Rev.\ C {\bf 90}, no. 4, 047301 (2014).


\bibitem{Isaka:2011kz} 
  M.~Isaka, M.~Kimura, A.~Dote and A.~Ohnishi,
  Phys.\ Rev.\ C {\bf 83}, 044323 (2011).


\bibitem{Isaka:2015xda} 
  M.~Isaka and M.~Kimura,
  Phys.\ Rev.\ C {\bf 92}, no. 4, 044326 (2015).

\bibitem{Homma:2015kia} 
  H.~Homma, M.~Isaka and M.~Kimura,
  Phys.\ Rev.\ C {\bf 91}, no. 1, 014314 (2015).

\bibitem{Isaka:2016apm} 
  M.~Isaka, Y.~Yamamoto and T.~A.~Rijken,
  Phys.\ Rev.\ C {\bf 94}, no. 4, 044310 (2016).

\bibitem{Isaka:2017nuc} 
  M.~Isaka, Y.~Yamamoto and T.~A.~Rijken,
  Phys.\ Rev.\ C {\bf 95}, no. 4, 044308 (2017).

\bibitem{Wirth:2014apa} 
  R.~Wirth, D.~Gazda, P.~Navr\'atil, A.~Calci, J.~Langhammer and R.~Roth,
  Phys.\ Rev.\ Lett.\  {\bf 113}, no. 19, 192502 (2014).

\bibitem{Rijken:2010zza} 
  T.~A.~Rijken, M.~M.~Nagels and Y.~Yamamoto,
  Nucl.\ Phys.\ A {\bf 835}, 160 (2010).

\bibitem{Rijken:2010zzb} 
  T.~A.~Rijken, M.~M.~Nagels and Y.~Yamamoto,
  Prog.\ Theor.\ Phys.\ Suppl.\  {\bf 185}, 14 (2010).

\bibitem{Yamamoto:2010zzn} 
  Y.~Yamamoto, T.~Motoba and T.~A.~Rijken,
  Prog.\ Theor.\ Phys.\ Suppl.\  {\bf 185}, 72 (2010).


\bibitem{KanadaEnyo:1995tb}
  Y.~Kanada-En'yo, H.~Horiuchi and A.~Ono,
  Phys.\ Rev.\  C {\bf 52}, 628  (1995).

\bibitem{KanadaEnyo:1995ir}
  Y.~Kanada-En'yo and H.~Horiuchi,
  Phys.\ Rev.\  C {\bf 52}, 647 (1995).

\bibitem{AMDsupp} 
Y. Kanada-En'yo and H. Horiuchi,
Prog. Theor. Phys. Suppl. {\bf 142},  205 (2001).

\bibitem{KanadaEn'yo:2012bj}
  Y.~Kanada-En'yo, M.~Kimura and A.~Ono,
 Prog. Theor. Exp. Phys. {\bf 2012}  01A202 (2012).


\bibitem{KanadaEn'yo:1998rf} 
  Y.~Kanada-En'yo,
  Phys.\ Rev.\ Lett.\  {\bf 81}, 5291 (1998).

\bibitem{Kanada-Enyo:2014qwn} 
  Y.~Kanada-En'yo and T.~Suhara,
  Phys.\ Rev.\ C {\bf 89}, no. 4, 044313 (2014).

\bibitem{Kanada-Enyo:2015uiy} 
  Y.~Kanada-En'yo, H.~Morita and F.~Kobayashi,
  Phys.\ Rev.\ C {\bf 91}, 054323 (2015).




\bibitem{Kanada-Enyo:2017ynk} 
  Y.~Kanada-En'yo,
  arXiv:1709.03375 [nucl-th].


\bibitem{Hill:1952jb} 
  D.~L.~Hill and J.~A.~Wheeler,
  Phys.\ Rev.\  {\bf 89}, 1102 (1953).

\bibitem{Griffin:1957zza} 
  J.~J.~Griffin and J.~A.~Wheeler,
  Phys.\ Rev.\  {\bf 108}, 311 (1957).


\bibitem{Kanada-Enyo:2018pxt} 
  Y.~Kanada-En'yo,
  arXiv:1801.01259 [nucl-th].



\bibitem{Brink66}
	D. M. Brink, {\it Proc. Int. School of Physics Enrico Fermi, Course 36}, Varenna, ed. C. Bloch (Academic Press, New York, 1966).

\bibitem{Kanada-Enyo:2016jnq} 
  Y.~Kanada-En'yo,
 Phys.\ Rev.\ C {\bf 94}, 024326 (2016).


\bibitem{Suhara:2014wua} 
  T.~Suhara and Y.~Kanada-En'yo,
  Phys.\ Rev.\ C {\bf 91}, no. 2, 024315 (2015).


\bibitem{Kanada-En'yo:2014oaa} 
  Y.~Kanada-En'yo and F.~Kobayashi,
  Phys.\ Rev.\ C {\bf 90}, no. 5, 054332 (2014).



\bibitem{Negele:1975zz} 
  J.~W.~Negele and D.~Vautherin,
  Phys.\ Rev.\ C {\bf 11}, 1031 (1975).

\bibitem{Kamimura:1988zz} 
  M.~Kamimura,
  Phys.\ Rev.\ A {\bf 38}, 621 (1988).

\bibitem{Hiyama:2003cu} 
  E.~Hiyama, Y.~Kino and M.~Kamimura,
  Prog.\ Part.\ Nucl.\ Phys.\  {\bf 51}, 223 (2003).

\bibitem{VOLKOV} A. B. Volkov, Nucl. Phys. {\bf 74}, 33 (1965).

\bibitem{LS}
 N. Yamaguchi, T. Kasahara, S. Nagata and Y. Akaishi,
  Prog. Theor. Phys. {\bf 62}, 1018  (1979);
 R. Tamagaki,  Prog. Theor. Phys. {\bf 39}, 91  (1968).


\bibitem{AjzenbergSelove:1990zh} 
  F.~Ajzenberg-Selove,
  Nucl.\ Phys.\ A {\bf 506}, 1 (1990).



\bibitem{Tilley:2002vg} 
  D.~R.~Tilley, C.~M.~Cheves, J.~L.~Godwin, G.~M.~Hale, H.~M.~Hofmann, J.~H.~Kelley, C.~G.~Sheu and H.~R.~Weller,
  Nucl.\ Phys.\ A {\bf 708}, 3 (2002).


\bibitem{Tilley:2004zz} 
  D.~R.~Tilley, J.~H.~Kelley, J.~L.~Godwin, D.~J.~Millener, J.~E.~Purcell, C.~G.~Sheu and H.~R.~Weller,
  Nucl.\ Phys.\ A {\bf 745}, 155 (2004).


\bibitem{Kelley:2012qua} 
  J.~H.~Kelley, E.~Kwan, J.~E.~Purcell, C.~G.~Sheu and H.~R.~Weller,
  Nucl.\ Phys.\ A {\bf 880}, 88 (2012).


\bibitem{Angeli13}
I.~Angeli, and K.~P.~Marinova,
Atom.~Data Nucl.~Data Tabl. {\bf 99}, 69 (2013). 





\bibitem{Juric:1973zq} 
  M.~Juri\u{c} {\it et al.},
  Nucl.\ Phys.\ B {\bf 52}, 1 (1973).

\bibitem{Davis:1992dt} 
  D.~H.~Davis,
  Nucl.\ Phys.\ A {\bf 547}, 369C (1992).


\bibitem{Davis:2005mb} 
  D.~H.~Davis,
  Nucl.\ Phys.\ A {\bf 754}, 3 (2005).

\bibitem{Ajimura:1998sy} 
  S.~Ajimura {\it et al.} [KEK-PS E336 Collaboration],
  Nucl.\ Phys.\ A {\bf 639}, 93 (1998).

\bibitem{Tang:2014atx} 
  L.~Tang {\it et al.} [HKS Collaboration],
  Phys.\ Rev.\ C {\bf 90},  034320 (2014).







\bibitem{Chrien:1990ag} 
  R.~E.~Chrien {\it et al.},
  Phys.\ Rev.\ C {\bf 41}, 1062 (1990).



\bibitem{Tanida:2000zs} 
  K.~Tanida {\it et al.},
  Phys.\ Rev.\ Lett.\  {\bf 86}, 1982 (2001).

\bibitem{Ajimura:2001na} 
  S.~Ajimura {\it et al.},
  Phys.\ Rev.\ Lett.\  {\bf 86}, 4255 (2001).
\bibitem{Kohri:2001nc} 
  H.~Kohri {\it et al.} [AGS-E929 Collaboration],
  Phys.\ Rev.\ C {\bf 65}, 034607 (2002).


\bibitem{Akikawa:2002tm} 
  H.~Akikawa {\it et al.},
  Phys.\ Rev.\ Lett.\  {\bf 88}, 082501 (2002).



\bibitem{Miura:2005mh} 
  Y.~Miura {\it et al.},
  Nucl.\ Phys.\ A {\bf 754}, 75 (2005).



\bibitem{Ukai:2006zp} 
  M.~Ukai {\it et al.} [E930'01 Collaboration],
  Phys.\ Rev.\ C {\bf 73}, 012501 (2006).

\bibitem{Ukai:2008aa} 
  M.~Ukai {\it et al.} [BNL E930 Collaboration],
  Phys.\ Rev.\ C {\bf 77}, 054315 (2008).



\bibitem{Ma:2010zzb} 
  Y.~Ma {\it et al.},
  Nucl.\ Phys.\ A {\bf 835}, 422 (2010).





\bibitem{Hosomi:2015fma} 
  K.~Hosomi {\it et al.},
  Prog. Theor. Exp. Phys.  {\bf 2015}, no. 8, 081D01 (2015).


\bibitem{Umeya-10Be} 
 A.~Umeya, private communication.

\bibitem{Motoba:2017lkk} 
  T.~Motoba,
  JPS Conf.\ Proc.\  {\bf 17}, 011003 (2017).

\bibitem{Umeya:2016bbt} 
  A.~Umeya and T.~Motoba,
  Nucl.\ Phys.\ A {\bf 954}, 242 (2016).











\end{thebibliography}
\end{document}